\documentclass[conference]{IEEEtran}
\makeatletter
\def\ps@headings{%
\def\@oddhead{\mbox{}\scriptsize\rightmark \hfil \thepage}%
\def\@evenhead{\scriptsize\thepage \hfil \leftmark\mbox{}}%
\def\@oddfoot{}%
\def\@evenfoot{}}
\makeatother
\usepackage[pdftex]{graphicx}
\usepackage[cmex10]{amsmath}
\usepackage{footnote}
\usepackage{color}
\usepackage{cite}
\usepackage{balance}
\usepackage{url}
\usepackage{amsfonts}
\usepackage{amssymb}
\usepackage{thm-restate}
\makesavenoteenv{tabular}
\usepackage{threeparttable}
\usepackage{multirow}
\usepackage{algorithmic}
\newcounter{MYtempeqncnt}

\usepackage[tight,footnotesize]{subfigure}

\begin{document}
%
\title{Wireless Multicast for Zoomable Video Streaming}


\author {Hui Wang, Mun Choon Chan and Wei Tsang Ooi\\
School of Computing, National University of Singapore \\
Email:\{wanghui,chanmc,ooiwt\}@comp.nus.edu.sg}

\maketitle

\begin{abstract}
Zoomable video streaming refers to a new class of interactive video
applications, where users can zoom into a video stream to view a selected region
of interest in higher resolutions and pan around to move the region of
interest.  The zoom and pan effects are typically achieved by breaking the
source video into a grid of independently decodable tiles.  Streaming the tiles
to a set of heterogeneous users using broadcast is challenging, as users have
different link rates and different regions of interest at different resolution
levels.  In this paper, we consider the following problem: given the subset of
tiles that each user requested, the link rate of each user, and the available
time slots, at which resolution should each tile be sent, to maximize the
overall video quality received by all users.  We design an efficient algorithm to
 solve the problem above, and evaluate the solution on a testbed using 10 mobile devices.
  Our method is able to achieve up to 12dB improvements over other heuristic methods. 
\end{abstract}

\section{Introduction}
\label{sec:intro}
Video streaming is a major Internet service that has been widely used to carry
both daily and major events (e.g., news, TV, sports, etc.).  With the
proliferation of mobile devices, streaming services continue to permeate
into our daily lives further.  Meanwhile, as technology evolves, videos with
increasingly higher resolution are becoming available (e.g., 8K UHD supports 33
Megapixels).  Due to screen size constraints (especially on mobile devices) and bandwidth constraints, however, video streaming playback is still limited in resolution.  As a result, high resolution videos are typically scaled down before transmission, leading to a loss in information.  

To address the mismatch of video resolution between the capture device and playback, zoomable video streaming has recently been proposed \cite{mavlankar2007optimal,mavlankar2007region,van2011spatial,quang2010supporting,shafiei2012jiku}.  A zoomable video supports zoom and pan as two new operations for a user to interact with the video.  In particular, a user is able to zoom into a selected region of interest (RoI) in the video, to view the RoI with higher resolution. The user essentially views the video through a viewport that defines a rectangular region in the high resolution video, from which the displayed video is cropped.  While zooming in, users can pan around by moving the viewport to view different regions in the video.  

In this paper, we are concerned with wireless multicasting of zoomable video streams, which can arise in scenarios such as interactive TV or live events such as broadcasting lectures in campus \cite{mavlankar2010interactive,halawa2011classx,shafiei2012jiku}, stage performances in concert, and sports in stadium (including eSports for spectating RTS games).  Multicast is a natural operation for transmitting these contents, as existing studies have reported that users tend to zoom into a small clusters of regions in the video~\cite{quang2010supporting} with substantial overlaps in their views.

In live zoomable video streaming system~\cite{shafiei2012jiku}, multiple resolution levels are available for each video stream.  For a given screen pixel size, the desired resolution level of a user depends on the size of the selected region of interest (RoI).  To stream
efficiently, the video is broken into a grid of small, independently decodable regions, each is termed as a \textit{tile} in this paper.  Instead of transmitting the whole frame, a minimum set of tiles covering the selected RoI with the desired resolution level is delivered.  
 
The problem that we consider in this paper is the following: given the available
time slots for video transmission and the selected RoI regions, how to
determine, for each tile, at which resolution level should it be multicasted to maximize the overall utility of all users?   There are two challenges in the aforementioned problem.  First, the scheme has to deal with changes in both RoI and the wireless channel that affects the supported link rates. Second, the solution has to be computational efficient and scalable (with respect to number of users/sessions, video qualities, link rate, and time horizon).  

In this work, we propose a novel and efficient algorithm to optimally solve this
zoomable multicast problem.  Our algorithm is inspired by several recent works
\cite{li2009scalable,yoon2012muvi} that look into the design of optimal algorithm
for video multicast allocation with a focus on heterogeneous link rates.  To
	evaluate our algorithm, we implemented the algorithm on a testbed that consists of the following key components: (i) mobile clients that support zoomable video functions, (ii) video server that supports streaming of zoomable video, and  (iii) a proxy that collects client RoI requests and wireless link conditions, runs the resource allocation algorithm, and multicasts the videos obtained from the server to the clients.

The major contributions of our work are as follows: (i) We model the zoomable video multicast problem as an optimization problem and develop an optimal algorithm that decides which resolution of which tile should be transmitted at which link rate.    The proposed optimal multicast improves the average video quality by up to 12dB, 6dB, and 3dB in terms of PSNR compared with three baseline schemes, \textit{adaptive unicast}, \textit{adaptive multicast}, and \textit{approximate multicast}, respectively.  
(ii) If we consider each tile as an individual video session, our proposed algorithm can be applied to the optimal allocation of multi-sessions adaptive video streaming as well, and has a lower, more practical, running time (grows linearly with the number of time slots) than the existing optimal allocation algorithms \cite{li2009scalable,yoon2012muvi}.   
(iii) We evaluate our solution on a wireless streaming testbed with up to 10 Android phones.

The rest of the paper is structured as follows. Section \ref{sec:relatework} discusses the related work. In Section \ref{sec:background}, we review the background of tiled zoomable streaming and mixed resolutions tiling scheme.  Section \ref{sec:problemDescription} states our maximization problem.  We present our optimal algorithm in Section \ref{sec:algorithm}.  The system implementations are detailed in Section \ref{sec:implementation} and performance evaluation results of our algorithm on Android platform are presented in Section \ref{sec:evaluation}.  The conclusion is made in Section \ref{sec:conclusion}.

\section{Related Work} \label{sec:relatework}
A tremendous amount of previous work aims at improving video multicast streaming system by dynamically adapting video data rates and multicast link rates.  In this section, we discuss only the most relevant work which can be broadly classified into three categories: adaptive video multicast, multicast link rate adaptation, and adaptive video with multicast link rate adaptation. 

\textbf{Adaptive video multicast.}  Many adaptive video multicast streaming approaches have been proposed to improve the performance of video streaming system \cite{chou2006rate,liu2003adaptive}.  Liu et al. \cite{liu2003adaptive} present an overview of existing studies and illustrate the advantages of adaptive streaming over non-adaptive streaming.  The problem of rate-adaptive optimized streaming is reduced to the error-cost optimized transmission problem in the following study~\cite{chou2006rate}.  This work also derives a fast practical algorithm to solve the formulated optimization problem.  Most of these works focus on adapting the video rate (quality) with the fixed low multicast link rate, which may under-utilize the networking resources.

\textbf{Multicast link rate adaptation.}  Recently, multicast link rate adaptation mechanisms have been suggested \cite{wongjoint,piamrat2009q,chandra2009dircast,sen2010scalable}.  Instead of using the basic rate, a relatively high broadcast rate is used for packet delivery, and FEC schemes are leveraged to protect the data from packet losses \cite{wongjoint,piamrat2009q}. 

Among these work, the most relevant works are DirCast~\cite{chandra2009dircast} and Medusa~\cite{sen2010scalable}.
DirCast multicasts packet at the link rate of the worst client for each access point (AP) and takes into account the rate anomaly problems. We adopt similar mechanisms.
Medusa prioritizes the frames according to their importance and transmits the less important frames at higher link rates.  By utilizing this frame level rate assignment heuristic, Medusa achieves higher video quality with limited resources.

\textbf{Adaptive video with multicast rate adaptation.}  To further improve streaming performance, the last category of research jointly adapts video data rate and multicast link rate.  This category is the most related to our work.  Deb et al.~\cite{wien2007real} investigate the utility optimization problem of scalable video multicast and prove that this problem is NP-Hard.  A greedy algorithm is then proposed to schedule the transmissions of layers and determine the corresponding modulation and coding scheme (MCS) assigned for each transmission.  Li et al.~\cite{li2009scalable} suggest a pseudo-polynomial algorithm with dynamic programming to solve the optimization problem.  Most recently, MuVi~\cite{yoon2012muvi} has been designed to investigate the optimal multicast scheduling problem for videos encoded with I, P, and B frames.  As the computational complexity of the suggested algorithm~\cite{li2009scalable} grows quadratically with the number of available time slots, it fails to efficiently solve the optimization problem with multiple multicast sessions.
To reduce the computational complexity especially for the case of multiple sessions, a fully polynomial time approximation algorithm is presented~\cite{li2010scalable}.  The approximation factor, however, linearly decreases with the number of multicast sessions.

Our approach falls under the category of adaptive video with multicast link rate adaptation. In contrast to previous work, we focus on a scenario where each user is interested in a subset of video tiles and user interests may partially overlapped.  Our algorithm can also be easily applied to the optimization problems with multiple sessions. 



\section{Background of mixed-resolutions tiled streaming} \label{sec:background}
In this section, we review the background of \textit{mixed resolutions tiling scheme}, which is proposed by Wang et al.~\cite{wang2014mixing}.  Moreover, to evaluate the perceptual quality of this scheme, the conducted psychophysical experiment is presented as well.


\subsection{Mixing Tile Resolutions in Tiled Video}
\begin{figure*}[!t]
\normalsize
\centering
\includegraphics[width=0.8\textwidth]{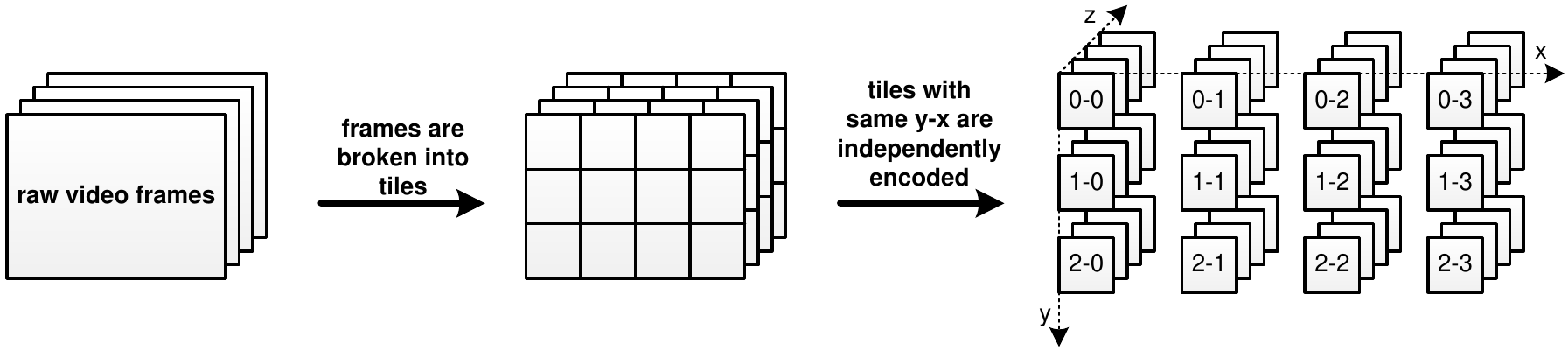}
\caption{Tile Video}
\label{fig:tilebroken}
\vspace*{4pt}
\end{figure*}

Zoomable video streaming is typically achieved using a technique called \textit{tiled streaming}, where video frames are broken into a grid of tiles (Figure \ref{fig:tilebroken}).  We can view the video as a three dimensional matrix of tiles.  Tiles at the same $y$-$x$ position in the matrix are temporally grouped and coded along $z$ axis.  The video is encoded into different resolutions to support zooming.  The zoom-out view corresponds to the lowest resolution.  As the user zooms in, a minimum set of tiles from the higher resolution video covering the RoI region is streamed.  The location of RoI can be changed by panning, while the resolution can be changed by zooming.  

The tiles in the same $y$-$x$ are decoded together by the zoomable player at the client side.  The tile groups with different $y$-$x$ positions can be decoded in parallel, each frame is formed by the uncompressed tiles with same $z$ position.   The frame will be displayed in the original order by the zoomable player when all the corresponding tiles are uncompressed.

In existing works~\cite{quang2010supporting,ngo2011adaptive,feng2011supporting}, at the server side, an original video is normally encoded into different versions (streams): frames of a low-resolution stream are constructed from a smaller number of tiles; and frames of higher-resolution streams are constructed from a larger number of tiles. At the client side, the number of tiles required to cover the physical screen resolution is fixed, therefore, the bandwidth consumption for each user will be mostly constant. Initially, a low resolution version of the video will be sent to users. When a user zooms into a RoI within the video, the server will first determine a suitable high-resolution stream based on the requested RoI size (zoom level).  It then selects tiles covering the requested RoI from this stream. This mechanism allows users to see their regions of interest in detail without consuming more bandwidth.

The afore-mentioned RoI cropping technique performs well in small scale networks by unicasting video stream.  In one of the use case we consider, the video stream is consumed by a large number of users within one location (e.g., in a concert hall or stadium).  To overcome the scalability issues with such a large number of users and RoI requests, wireless multicast scheme is employed.   
When the RoI regions from multiple users partially overlap, tiles from the overlapped regions could be potentially multicasted to all interested users to save bandwidth consumption.  In zoomable video, different users, however, may have different zoom levels (i.e., different RoI sizes) and will need tiles from different versions encoded at different resolutions, which prevents the potential benefits of wireless multicast. 


Instead of fixing tile size, using a fixed number of tiles to encode and decode videos could be more effective.  At the server side, an original video will be encoded into different resolution versions, but all versions consist of the same number of tiles.   The same amount of tiles is required at the client side to decode each video frame.  Within a frame, however, different tiles could come from different resolution streams.  If a tile comes from a stream with resolution lower/higher than requested level, it will be scaled up/down accordingly.  In zoomable video, when a user zooms into a region of interest (RoI) within the video, the server will first determine the tiles covering this RoI, and then associate each tile with an appropriate stream version, depending on their popularity and the resource constraints.  

The proposed \textit{mixed resolutions tiling} scheme has the following two essential advantages in tiled video streaming.  First, benefiting from the scaling up/down operations for each tile, the multicast transmissions are considerably reduced.  Next, by intelligently allocating resolution version to each tile, the mixing resolutions approach may considerably reduce bandwidth consumption without impairing much perceived video quality.  For instance, the popular regions/tiles requested by many users could come from high-resolution streams; while tiles requested by one or few users could come from a low-resolution stream under limited bandwidth condition. 

Although this proposed scheme saves bandwidth, the impairment to the perceived quality is still unclear.  Thus, to understand if, and at what thresholds, users could notice and /or accept the difference between original video and tiled video with mixed resolutions, we conducted a psychophysical study with 50 participants, which is presented in the remainder of this section.

\subsection{Perceptual Quality Assessment} \label{sec:qualityassessment}
Using the \textit{method of limits} from psychophysics~\cite{gescheider1997psychophysics}, we measure two perceptual thresholds -- Just Noticeable Difference (JND) and Just Unacceptable Difference (JUD) -- to understand the user perception about the quality of mixed-resolution tiled video.  The two identified difference thresholds partition the quality degradation level (introduced by mixing tile resolutions) into the following three intervals: without noticeable quality degradation, with noticeable (but acceptable) quality degradation, and with unacceptable quality degradation.

\begin{table} [ht]
\renewcommand{\arraystretch}{1.2}
\caption{The number of pixels in each frame and each tile at different resolution levels.}
\label{tab:resolutionsize1}
\centering
{
\begin{threeparttable} [b]
\begin{tabular}{ | c || c | c | c | } 
\cline{1-4}
{level} & frame & 16$\times$9 tiles &  80$\times$45 tiles \\ \cline{1-4}
5 &1920$\times$1080 & 120$\times$120 & 24$\times$24 \\ \cline{1-4}
4 &1600$\times$900  & 100$\times$100 & 20$\times$20 \\ \cline{1-4}
3 &1280$\times$720 & 80$\times$80 & 16$\times$16 \\ \cline{1-4}
2 &960$\times$540 & 60$\times$60 & 12$\times$12 \\ \cline{1-4}
1 &640$\times$360 & 40$\times$40 & 8$\times$8 \\ \cline{1-4}
\end{tabular} 
\end{threeparttable} }
\end{table} 

\subsubsection{Setup}
The experiments assess the quality of mixed-resolution tiled video using three standard HD (1920$\times$1080p) test video files, \textit{Crowd-Run} (dense motion, 50fps), \textit{Old-Town-Cross} (medium motion, 50fps), and \textit{Rush-Hour} (low motion, 25fps)\footnote{Available at \url{http://media.xiph.org/video/derf/}}.  We have five resolution levels for each video file, these levels are labeled from 5 to 1 (Table \ref{tab:resolutionsize1}).   The pixels of the original video frame at five resolution levels are: 1920$\times$1080, 1600$\times$900, 1280$\times$720, 960$\times$540, and 640$\times$360.   

\begin{figure*}[!t]
\normalsize
\centering
\subfigure[Tiles from HD stream]{\includegraphics[width=0.3\textwidth]{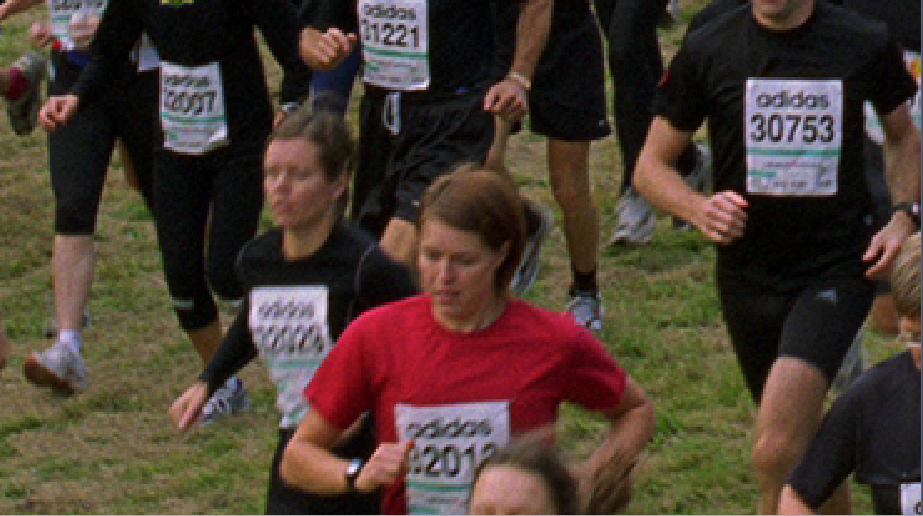}
\label{fig:highres1}}
\hfil
\subfigure[Mixing tiles from stream levels 5 and 3]{\includegraphics[width=0.3\textwidth]{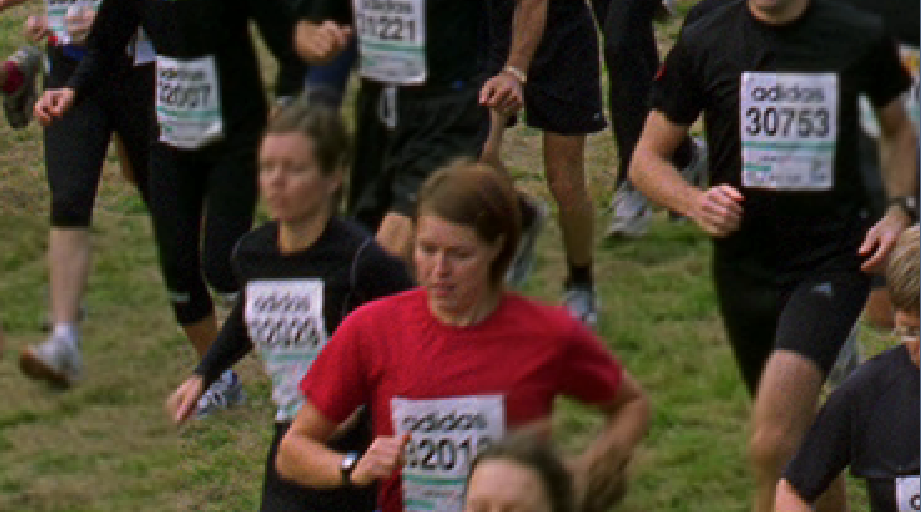}
\label{fig:medium_high1}}
\hfil
\subfigure[Mixing tiles from stream levels 5 and 1]{\includegraphics[width=0.3\textwidth]{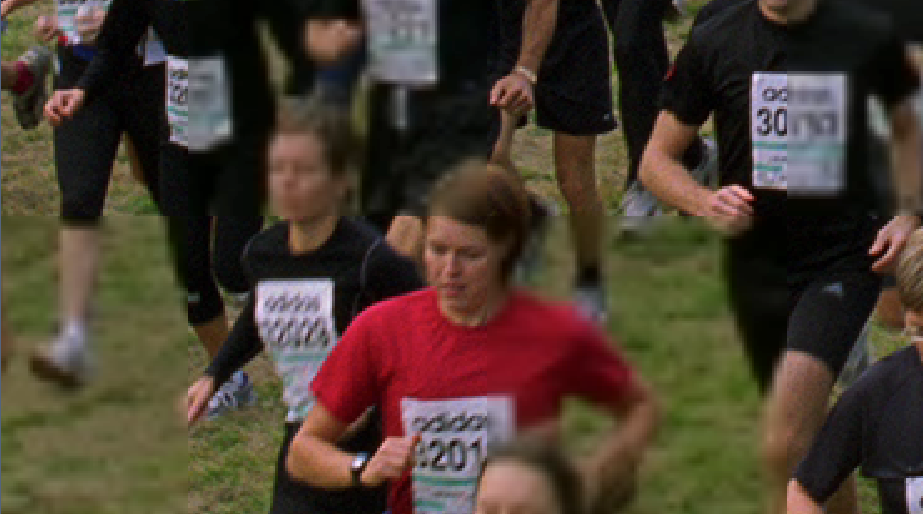}
\label{fig:low_high1}}
\caption{Mixing tile resolutions of {Crowd-Run}}
\label{fig:mixedResVideoCrowdRun}
\end{figure*}

\begin{figure*}[!t]
\normalsize
\centering
\subfigure[Tiles from HD stream]{\includegraphics[width=0.3\textwidth]{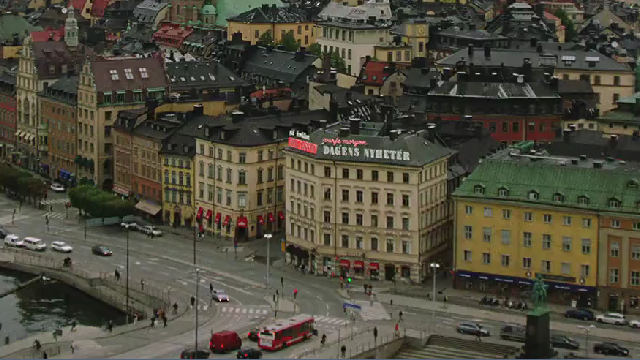}
\label{fig:highres2}}
\hfil
\subfigure[Mixing tiles from stream levels 5 and 3]{\includegraphics[width=0.3\textwidth]{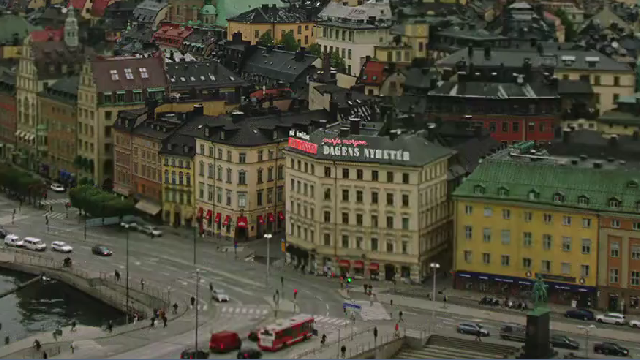}
\label{fig:medium_high2}}
\hfil
\subfigure[Mixing tiles from stream levels 5 and 1]{\includegraphics[width=0.3\textwidth]{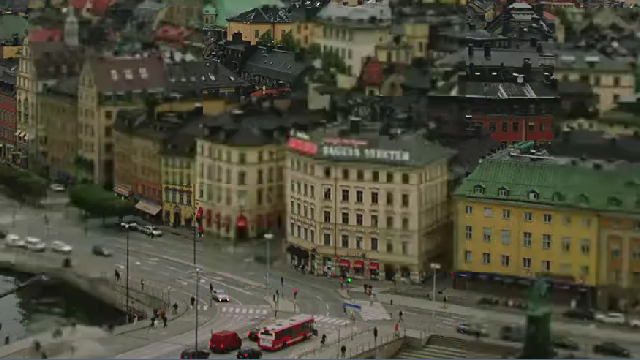}
\label{fig:low_high2}}
\caption{Mixing tile resolutions of {Old-Town-Cross}}
\label{fig:mixedResVideoOldTownCross}
\end{figure*}

\begin{figure*}[!t]
\normalsize
\centering
\subfigure[Tiles from HD stream]{\includegraphics[width=0.3\textwidth]{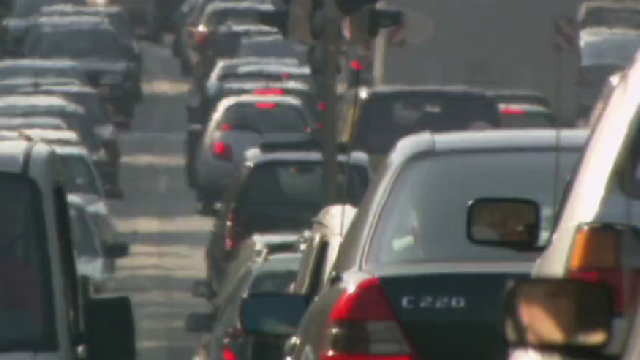}
\label{fig:highres3}}
\hfil
\subfigure[Mixing tiles from stream levels 5 and 3]{\includegraphics[width=0.3\textwidth]{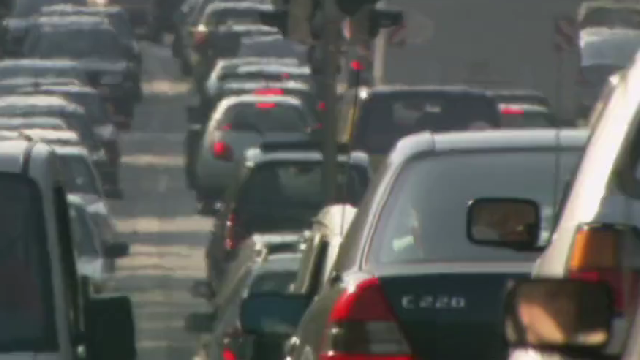}
\label{fig:medium_high3}}
\hfil
\subfigure[Mixing tiles from stream levels 5 and 1]{\includegraphics[width=0.3\textwidth]{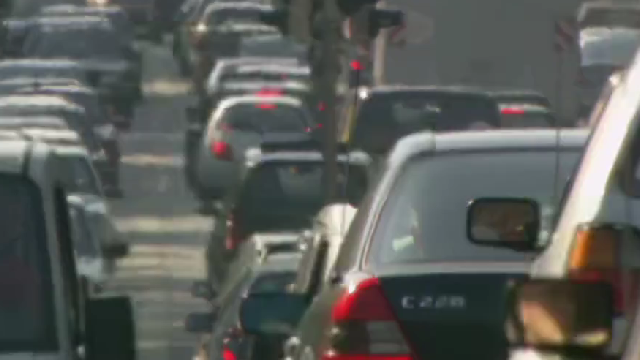}
\label{fig:low_high3}}
\caption{Mixing tile resolutions of {Rush-Hour}}
\label{fig:mixedResVideoRushhour}
\end{figure*}

In the experiments, we construct mixed-resolution tiled video by mixing two resolution levels, where the higher resolution level is denoted as $R_H$ and the lower resolution level is denoted as $R_L$.  Specifically, given a pair of $R_H$ and $R_L$, we randomly allocate resolution level $R_H$ or $R_L$ to each tile with equal probability.  For any particular pair of $R_H$ and $R_L$, we restrict the range of $R_H$ as $3 \leq R_H \leq 5$ and the range of $R_L$ as $1 \leq R_L \leq R_H$.   Figures~\ref{fig:mixedResVideoCrowdRun},~\ref{fig:mixedResVideoOldTownCross}, and~\ref{fig:mixedResVideoRushhour} show the screenshots of mixed-resolution tiled video.  

Since the aspect ratio of the test HD video frame sequences is 16:9, we break the video frames into 16$\times$9 tiles by default.  As a result, each tile size (view region size) is $\frac{1}{16 \times 9}$ of the entire view region.  To evaluate the impact of tile size, in addition to the default configuration, we generate another set of videos where each video frame is broken into 80$\times$45 tiles.  The number of pixels for a tile at each resolution level is shown in Table \ref{tab:resolutionsize1}.

\subsubsection{Procedures}
Fifty adult participants were invited to participate in our assessment, primarily graduate students and research staffs from National University of Singapore.  The sample consisted of 16 women and 34 men; all had normal vision.  They were asked to watch the mixed-resolution tiled videos online\footnote{Online website is available at: \\* \url{http://liubei.ddns.comp.nus.edu.sg/resMix}} using a monitor with full HD display resolution.

For configurations with 16$\times$9 tiles, we vary the high resolution level $R_H$ from $5$ to $3$, 9 stimuli series are generated over three test videos.  For configurations with 80$\times$45 tiles, we generate stimuli series with $R_H=5$.  As a result, we have 12 stimuli series in total, which are shuffled in a random order and played.  
 
 \begin{figure}[ht]
 \centering
 \subfigure[Ascending stimuli series]{\includegraphics[width=0.4\textwidth]{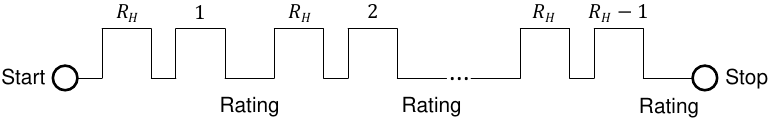}
 \label{fig:ascending}}
 \hfil
 \subfigure[Descending stimuli series]{\includegraphics[width=0.4\textwidth]{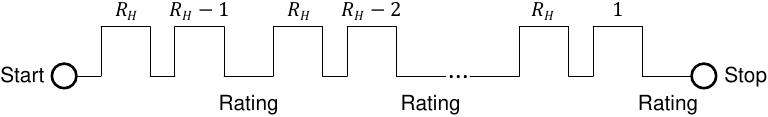}
 \label{fig:descending}}
 \caption{Experiment procedure.  The video is composed by tiles with resolution level $R_H$ and $R_L$.  The numbers above represent the value of $R_L$, the first video in each pair is a standard tiled video where $R_L=R_H$, and the second video is a mixed-resolution tiled video.}
 \label{fig:procedure}
 \end{figure}

 For each series, the stimuli is randomly manipulated in either an ascending or a descending order, the procedures are depicted in Figure \ref{fig:procedure}.  In a stimuli series, we fix the high resolution level $R_H$ and vary the low resolution level $R_L$.  As shown in the figure, each pair presents a standard video where $R_L=R_H$ and a mixed-resolution tiled video.  After watching the videos in a pair ($10s$ per video), the participant is asked to rate the level of the difference between two videos.  In particular, two questions are asked: (i) \textit{is the quality difference noticeable} and (ii) \textit{is the quality difference unacceptable}.   In the case of ascending series, we increase $R_{L}$ from 1.  On each successive trial, we increase $R_L$ by 1 until the participant eventually reports the difference is unnoticeable or $R_L=R_H-1$.  If the series is descending, the stimuli operates in an opposite direction.  We start from $R_L=R_H-1$ and gradually decrease $R_L$ until the participant reports the difference is unacceptable or $R_L=1$.  

 Using the above procedure, the obtained results fall into the following three categories: (i) The noticeable difference threshold and unacceptable difference threshold are both detected;  (ii) Only the noticeable difference threshold is detected; and (iii) Neither noticeable difference threshold nor unacceptable difference threshold can be detected.  Assuming that we have detected the noticeable difference threshold and unacceptable threshold, denoted by $T_{ND}$ and $T_{UD}$, respectively, then according to \textit{the method of limits}~\cite{gescheider1997psychophysics}, we estimate the Just Noticeable Difference threshold as $\big(T_{ND}+(T_{ND}+1)\big)/2=T_{ND}+0.5$.  Similarly, we express Just Unacceptable Difference threshold as $\big(T_{UD} + (T_{UD}+1)\big)/2=T_{UD}+0.5$.  For the cases where we failed to detect the difference threshold, we set the corresponding Just Noticeable/Unacceptable Difference threshold to $0$.

 \begin{figure*}[!t]
 \centering
 \begin{minipage}[b]{0.46\linewidth}
 \centering
 \includegraphics[width=\textwidth]{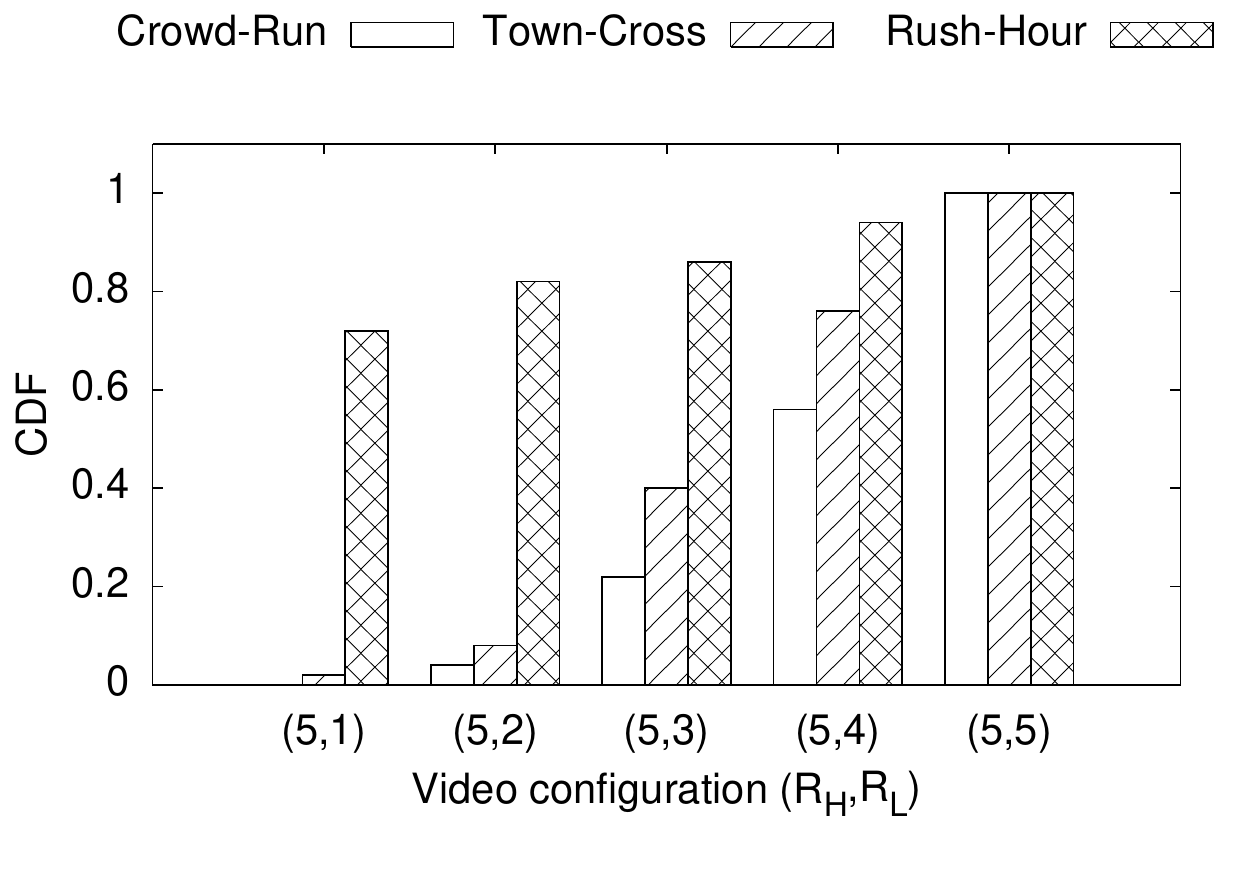}
 \caption{CDF distribution of participants that cannot notice any difference between mixed-resolution tiled video (5, $R_L$) and standard HD tiled video (5, 5).}
 \label{fig:cdf_unnoticeable}
 \end{minipage}
 \hspace{0.5cm}
 \begin{minipage}[b]{0.46\linewidth}
 \centering
 \includegraphics[width=\textwidth]{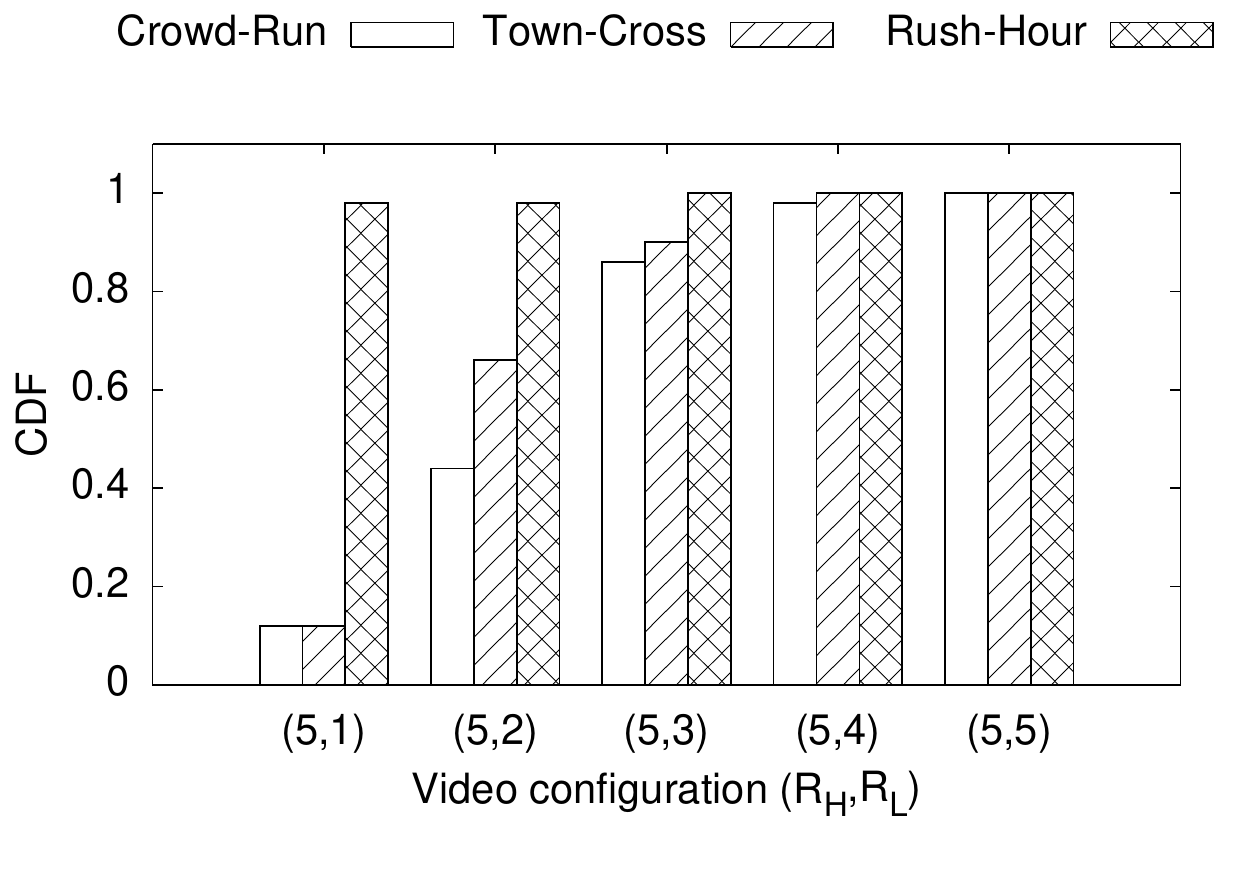}
 \caption{CDF distribution of participants that accept the quality difference between mixed-resolution tiled video (5, $R_L$) and standard HD tiled video (5, 5).}
 \label{fig:cdf_accept}
 \end{minipage}
 \vspace*{4pt}
 \end{figure*}

\subsubsection{Results}
We first examine the configuration with 16$\times$9 tiles.  Figure \ref{fig:cdf_unnoticeable} depicts the CDF distribution of participants that cannot notice any difference between mixed-resolution tiled video (5, $R_L$) and standard tiled HD video (5, 5). The CDF distribution of participants that accept the quality difference is present in Figure \ref{fig:cdf_accept}.    The average measured thresholds of Just Noticeable Difference and Just Unacceptable Difference for $R_H$ in the range from 5 to 3 are shown in Table \ref{tab:16and9JND} and Table \ref{tab:16and9JUD}, respectively.  
\begin{table} [ht]
\renewcommand{\arraystretch}{1.2}
\caption{The average Just Noticeable Difference threshold (number within parenthesis is the 95\% Confidence Interval value).}
\label{tab:16and9JND}
\centering
{
\begin{threeparttable} [b]
\begin{tabular}{ | c || c | c | c | } 
\cline{1-4}
$R_H$ & Crowd-Run & Old-Town-Cross &  Rush-Hour \\ \cline{1-4}
5 & 3.68 ($\pm$0.52) & 3.25 ($\pm$0.47) & 0.81 ($\pm$0.23) \\ \cline{1-4}
4 & 2.74 ($\pm$0.39) & 2.31 ($\pm$0.34) & 0.24 ($\pm$0.10) \\ \cline{1-4}
3 & 2.09 ($\pm$0.30) & 1.73 ($\pm$0.26) & 0.11 ($\pm$0.06) \\ \cline{1-4}
\end{tabular} 
\end{threeparttable} }
\end{table} 

\begin{table} [ht]
\renewcommand{\arraystretch}{1.2}
\caption{The average Just Unacceptable Difference threshold (number within parenthesis is the 95\% Confidence Interval value).}
\label{tab:16and9JUD}
\centering
{
\begin{threeparttable} [b]
\begin{tabular}{ | c || c | c | c | } 
\cline{1-4}
$R_H$ & Crowd-Run & Old-Town-Cross &  Rush-Hour \\ \cline{1-4}
5 & 2.03 ($\pm$0.31) & 1.76 ($\pm$0.27) & $0(0)$ \\ \cline{1-4}
4 & 1.64 ($\pm$0.26) & 1.28 ($\pm$0.21) & $0(0)$ \\ \cline{1-4}
3 & 1.28 ($\pm$0.21) & 0.69 ($\pm$0.14) & $0(0)$ \\ \cline{1-4}
\end{tabular} 
\end{threeparttable} }
\end{table} 

From the results, we can draw the following observations.

\textbf{Feasibility of Mixing Tile Resolutions.}  The measured thresholds confirm the feasibility of mixed-resolution tiled video.  The CDF distribution from Figure~\ref{fig:cdf_unnoticeable} implies that we can mix tiles with resolution levels 5 and 4 without being noticed in most cases.   Further, the depicted result from Figure~\ref{fig:cdf_accept} indicates that more than 85\% participants accept the quality difference with configurations where $3 \leq R_L \leq R_H=5$; under these configurations, up to 30\% bandwidth can be saved by mixing tile resolutions.  When we construct video from tiles at resolution level 5 and 2, almost all participants noticed the difference for video \textit{Crowd-Run} and \textit{Old-Town-Cross}.  40\% to 65\% of the participants, however, still accept the quality difference.


\textbf{Impact of Content.}  With the same configuration, the results from Tables~\ref{tab:16and9JND} and~\ref{tab:16and9JUD} show a great disparity in the measured average JND and JUD across three test videos.  Overall, video \textit{Crowd-Run}, which has the highest amount of motion among the three test videos,  is most sensitive to the resolution mixing, as the highest average threshold and the greatest variation are detected.  Interestingly, video \textit{Rush-Hour}, which has the lowest amount of motion among the three test videos,  performs remarkably different from others. It is difficult to notice the quality difference between the mixed-resolution tiled video and the standard version, thus the average measured thresholds and the variations are much smaller compared with other test videos.

\textbf{Gap between JND and JUD Thresholds.}  For many cases, although participants could notice the difference, it is still acceptable.  Generally, a greater gap value indicates a higher video quality tolerance degree when the quality difference is noticeable.  From the Tables~\ref{tab:16and9JND} and \ref{tab:16and9JUD}, we observe a significant gap between the average measured JND and JUD thresholds, especially for $R_H=5$.   In particular, the average gap quantities for video \textit{Crowd-Run} and \textit{Old-Town-Cross} with $R_H=5$ are 1.65 and 1.49, respectively.  As the tolerance space is reduced with smaller $R_H$ value, the quantity of the threshold gap between JND and JUD will be reduced as well, as can be seen in both tables.
\begin{table} [ht]
\renewcommand{\arraystretch}{1.2}
\caption{The average Just Noticeable Difference threshold where $R_H=5$ (number within parenthesis is the 95\% Confidence Interval value).}
\label{tab:jnd16and9and80and45}
\centering
{
\begin{threeparttable} [b]
\begin{tabular}{ | c || c | c | c | } 
\cline{1-4}
& Crowd-Run & Old-Town-Cross &  Rush-Hour \\ \cline{1-4}
16$\times$9 &  3.68 ($\pm$0.52) & 3.25 ($\pm$0.47) & 0.81 ($\pm$0.23) \\ \cline{1-4}
80$\times$45 & 3.30 ($\pm$0.48) & 3.04 ($\pm$0.44) & 0.76 ($\pm$0.20) \\ \cline{1-4}
\end{tabular} 
\end{threeparttable} }
\end{table} 

\begin{table} [ht]
\renewcommand{\arraystretch}{1.2}
\caption{The average Just Unacceptable Difference threshold where $R_H=5$ (number within parenthesis is the 95\% Confidence Interval value).}
\label{tab:jud16and9and80and45}
\centering
{
\begin{threeparttable} [b]
\begin{tabular}{ | c || c | c | c | } 
\cline{1-4}
& Crowd-Run & Old-Town-Cross &  Rush-Hour \\ \cline{1-4}
16$\times$9 & 2.03 ($\pm$0.31) & 1.76 ($\pm$0.27) & $0(0)$ \\ \cline{1-4}
80$\times$45 & 1.76 ($\pm$0.29) & 1.63 ($\pm$0.25) & $0(0)$ \\ \cline{1-4}
\end{tabular} 
\end{threeparttable} }
\end{table} 

\textbf{Impact of Tile Size.}  The comparison between the configurations with 16$\times$9 tiles and 80$\times$45 tiles is present in Tables \ref{tab:jnd16and9and80and45} and \ref{tab:jud16and9and80and45}.  The threshold values with 80$\times$45 tiles is slightly smaller than the corresponding threshold values with 16$\times$9 tiles, which indicates that the quality degradation introduced by mixing resolutions is slightly less obvious for the finer-grained tile size (80$\times$45) compared with the coarse-grained tile size (16$\times$9).  The finer-grained tiles, however, are generally less efficient in terms of encoding and transmission bandwidth.  Therefore, we need to balance the trade-off between the video quality and the efficiency to obtain an appropriate configuration.   

\subsection{Summary} \label{sec:assessmentSummary}
The subjective assessment demonstrated that in most cases, the perceptual quality loss of mixing resolutions in tiled video is insignificant, as long as the variance of mixed resolution levels is low.  From the evaluation results, we can draw the following two important observations:
\begin{itemize}
\item In most cases, tiles from 1920$\times$1080p stream and 1600$\times$900p stream could be mixed together without being noticed;
\item Even when participants could notice quality degradation in videos combined with tiles from 1920$\times$1080p stream and tiles from 960$\times$540p stream, greater than 80\% of participants still accept the quality difference for low and medium motion videos; and more than 40\% of participants accept the quality difference for the dense motion video.
\end{itemize}

This section confirms the feasibility of \textit{mixed resolution tiling} scheme, which will be applied to wireless multicast of tiled video streams in the rest of this paper.  Instead of randomly mixing resolutions of tiles, we are looking into how to optimally allocate resolution versions to each tile to better utilize the wireless bandwidth and improve overall utilities of users.


\section{Problem Definition} \label{sec:problemDescription}
We now describe an optimization problem to determine which tile should be sent at which resolution and at which link rate, given the wireless network constraint.  Let $T$ be the number of slots available on average for the delivery of a single frame, where a slot refers to a minimum transmission time unit in $802.11$ network (e.g., $9\mu s$ in $802.11a$).  The wireless network supports $N_r$ different link rates.  Let $n$ be the number of users in our system; the physical link rates of these $n$ clients are: $r_1, r_2, \ldots, r_n$.  Without loss of generality, we assume that link rate $r_i$ is a non-decreasing function of index $i$.   

We generate $M$ resolution versions (or levels) for each frame, and every frame is broken into $N_g$ view regions, each view region is termed as a tile (or grid).  Instead of using the y-x notation in Figure \ref{fig:tilebroken}, we simply number the tiles $1, 2, \ldots, N_g$ when we discuss the algorithm.  A tile is considered as a logical entity -- when transmitted, a tile has to have a specific resolution level.  A tile $g$ with resolution level $m$ ($1 \leq m \leq M$) is denoted by $g_m$, the size of which is $s_g^m$.  The sequence of $s_g^1, s_g^2, \ldots, s_g^M$ is strictly increasing.

The set of tiles in the RoI of user $i$ is denoted as $\mathbf{G}(i)$.  Let $R_i$ be the request resolution level from user $i$.  With restricted bandwidth condition, we may not be able to satisfy all the user requests.  As a result, some tiles may be streamed with resolution levels lower than the desired resolution level ($R_i$).  To avoid significant perceptual quality loss introduced by downgrading tile resolution levels, for user $i$, we have a lower bound $L_i$ of the tile resolution levels, which is guaranteed to be satisfied.  More specifically, for every tile in $\mathbf{G}(i)$, the resolution level to be decoded (the highest received level) by user $i$ should be at least $L_i$.   

Receiving $g_m$ at user $i$ yields utility $u_{g,i}^m$, which follows the rules below:
\begin{itemize}
\item If $g \notin \mathbf{G}(i)$, then $u_{g,i}^m = 0$ (for all $1 \leq m \leq M)$;
\item If $g \in \mathbf{G}(i)$ and $m < L_i$, we have $u_{g,i}^m = -\infty$;
\item If $g \in \mathbf{G}(i)$ and $L_i \leq m < m' \leq R_i$, we have $u_{g,i}^m < u_{g,i}^{m'}$;
\item If $g \in \mathbf{G}(i)$ and $m > R_i$, we have $u_{g,i}^m = u_{g,i}^{R_i}$.
\end{itemize}

For simplicity, we use a tile size-based utility assignment mechanism.  In particular, $u_{g,i}^{R_i}$ is the maximum achievable utility at user $i$ by receiving tile $g$, the utility assignments of receiving other levels are proportional to the corresponding tile sizes.  The utility function, however, can be any general function (e.g., the PSNR of tiles) subject to the above rules.

Given the RoI selection and the corresponding utility assignment of tiles with each resolution level  at each user, the objective is to maximize the total utility received by all users subject to the total transmission slot constraint.

Lastly, we discuss the parameter settings for the average available time slots $T$ and the tile size with a specific resolution level.  All pixels belonging to the  same tile across different frames will be encoded as a group of picture (GOP).  Due to the dependency in a GOP, if we pick a resolution level $m$ for a tile, we have to transmit this tile at the same resolution $m$ for all frames within the same GOP.  In our model, we therefore model $s_g^m$ as the average tile size in a GOP and model the average number of time slots needed per frame as $T$.  In the implementation, however, the time slots allocated to frames in a GOP is proportionally distributed according to the actual frame sizes, as there is a considerable diversity in the sizes of I, B, and P frames.

\begin{table}[ht]
\renewcommand{\arraystretch}{1.3}
\caption{Key Notations in the Algorithm}
\label{tab:notation}
\centering
{\fontsize{8}{9} \selectfont
\begin{tabular}{|c | l |}
\hline
Notation & Definition\\
\hline
$T$ & Average available time slots for delivering one frame\\
$n$ & Number of users (or clients)\\
$r_i$ & The link rate of user $i$ \\
$N_r$ & Number of different link rate levels \\
$M$ & The number of available resolution levels \\
$\mathbf{G}(i)$ & The set of tiles in the RoI of user $i$ \\
$R_i$ & Resolution level requested by user $i$ ($R_i \geq 1$)\\
$L_i$ & Resolution level guaranteed to be satisfied for user $i$ ($L_i \geq 1$)  \\
$N_g$ & Total number of tiles (or grids) \\
$g_m$ & Tile $g$ at resolution level $m$ \\
$s_g^m$ & The size (in bytes) of tile $g$ at resolution level $m$ \\
$u_{g, i}^m$ & The utility of tile $g$ at resolution level $m$ for user $i$ \\
$\mathcal{M}(g, i)$ & The highest received resolution level of tile $g$ for user $i$ \\
\hline
\end{tabular}
}
\end{table}
The key notations used in this paper are summarized in Table~\ref{tab:notation}.  

\section{Optimal Broadcast Algorithm} \label{sec:algorithm}
This section presents a dynamic programming algorithm to solve the utility maximization problem defined in the previous section.  The solution consists of three major components: (i) an algorithm that determines an appropriate quality lower bound for each user; (ii) an optimal algorithm for determining the link rate and resolution level of a single tile; and (iii) an efficient algorithm for determining the link rate and resolution level over multiple tiles. 

\subsection{Adaptive Utility Assignment} \label{sec:aUtilityAssign}
The mixture of resolution levels could result in two potential issues when the available bandwidth is insufficient to meet the requirements from all users.  First, as discussed in Section~\ref{sec:background}, the significant disparity of resolution levels between tiles for a user may severely impair the visual perception.  Next, the utility-oriented optimization algorithm could result in severe unfairness.  To address these issues, we suggest an algorithm to adaptively tune the lower bound $L_i (1 \leq i \leq n)$ of resolution level that is guaranteed to be satisfied.

Recall that $R_i$ is the requested resolution level from user $i$, and $L_i$ is the resolution level guaranteed to be satisfied for user $i$ among its interested tiles.  Given $R_i$ and $L_i$, the rules for utility assignment are specified in Section \ref{sec:problemDescription}.  It is clear that when all requests of users are satisfied, we have $L_i \geq R_i$ for any $1 \leq i \leq n$, and the overall utility is optimal.  Hence, we set $L_i = R_i$ at the beginning, then we validate the feasibility of current configuration for $L_i$ and adapt accordingly.   

We define an indicator variable $x_{g,i}^m$, which takes the value of 1 if resolution level $m$ of tile $g$ is transmitted at link rate $r_i$, and 0 otherwise.  Let $\mathcal{M}(g, i)$ be the maximum  resolution level of tile $g$ to be received by user $i$.  Since user $i$ can only receive the transmissions with link rates not higher than $r_i$, the expression of $\mathcal{M}(g, i)$ can be written as $\mathcal{M}(g, i) = \max {\left \{ m \vert x_{g,i'}^m = 1\text{ and }1 \leq i' \leq i  \right\} }$.   Now we can formulate the feasibility validation problem as
\begin{align}
\label{eq:receiveTile}
 & \sum\limits_{g=1}^{N_g}{\sum\limits_{i=1}^{n}{ u_{g,i}^{\mathcal{M}(g, i)} }} \geq 0 ,   \\
\label{eq:timeslotLimit}
 \text{subject to } & \sum\limits_{g=1}^{N_g} { \sum\limits_{m=1}^{M} { \sum\limits_{i=1}^{n} \bigg( x_{g,i}^m \cdot \left \lceil{\frac{s_g^m}{r_i}}\right \rceil \bigg)  }} \leq T , 
\end{align}
where $u_{g,i}^{\mathcal{M}(g, i)} = -\infty$ if $g \in \mathbf{G}(i)$ and $\mathcal{M}(g, i) < L_i$; the unit of expression $s_g^m/r_i$ is a $802.11$ time slot.  To obtain an appropriate setting of $L_i$, we keep decreasing $L_i$ by $1$ for all $i$ until Inequality (\ref{eq:receiveTile}) is feasible subject to time limit constraint (\ref{eq:timeslotLimit}).  

To solve the feasibility problem defined above, we first independently calculate the minimum required time slots for every tile $g$ $(1 \leq g \leq N_g)$ and then simply integrate the required time slots across all $N_g$ tiles.   The total required slots should be less or equal than $T$, if the current lower bound requirement ($L_i$) is achievable.   The following paragraph presents an algorithm to calculate the minimum required time slot for any single tile $g$.

For user $i$, the lower bound requirement of resolution level $L_i$ can be satisfied by either transmitting at link rate $r_i$ or at lower link rate $r_{i'}$, where $1 \leq i' \leq i$.  Define $\Re_g(i, l)$ as the minimum required time slots satisfying non-negative utility requirement with users up to $i$ and with resolution level $l$ has not been satisfied from users with indexes larger than $i$.  The recursive equation for $\Re_g(i, l)$ can be written as 

{\fontsize{8}{9} \selectfont 
\begin{equation}
\label{eq:minimumTimeSlot}
\Re_g(i, l) = \left\{ 
  \begin{array}{l l}
    \min \left \{ \Re_g(i-1, H), \Re_g(i-1, 0)+\left \lceil{\frac{s_g^m}{r_i}}\right \rceil\right \}, &\text{if $g \in \mathbf{G}(i)$;}      \\
    \Re_g(i-1, l), &\text{if $g \notin \mathbf{G}(i)$,}
  \end{array} \right.
\end{equation}
}
where $H = \max \{l, L_i \}$.  The minimum time slots required for delivering tile $g$ while satisfying the quality lower bound is $\Re_g(n, 0)$, which could be easily calculated by leveraging recursion (\ref{eq:minimumTimeSlot}).
Now we are able to simplify the feasibility validation problem to $\sum_{g=1}^{N_g} \Re_g(n, 0) \leq T$.

\subsection{Optimal Allocation for a Single Tile}
\label{sec:singleTileSchedule}
For ease of analysis, we begin with designing an optimal resource allocation algorithm for a single tile.  We denote this particular tile as $g$.   The optimal allocation approach determines \emph{the resolution levels of tile $g$ to be transmitted} and \emph{the link rate for each transmission}.

\subsubsection{Optimal Allocation Algorithm}
Let $t$ ($0 \leq t \leq T$) be the total slots available for the transmissions of tile $g$.  The utility optimization problem can be formulated as
\begin{align}
 \text{maximize } & \sum\limits_{i=1}^{n}u_{g,i}^{\mathcal{M}(g,i)},  \nonumber \\
 \text{subject to } & \sum\limits_{m=1}^{M} { \sum\limits_{i=1}^{n} \bigg( x_{g,i}^m \cdot \left \lceil{\frac{s_g^m}{r_i}}\right \rceil \bigg)  } \leq t.
\end{align}

As we assume that the users with higher link rate can receive all transmissions at lower rates, we have the following important observation: \emph{for any tile, a higher resolution version is always transmitted with higher link rate}.  By utilizing this observation, we have the following definition of the maximum utility function.   For tile $g$, define $\mathcal{U}_g(i, m, t)$ as the optimal utility with users $u_1, u_2, \ldots, u_i$, with resolution levels up to $m$, and within transmission time limit $t$.  

Every state $\mathcal{U}_g(i, m, t)$ falls into category of either \emph{user $i$ is not interested in tile $g$} or \emph{user $i$ is interested in tile $g$}.  If user $i$ is not interested in tile $g$ ($g \not \in \mathbf{G}(i)$), the state transition equation could be simply written as
\begin{equation}
\label{eq:notInterested}
\mathcal{U}_g(i, m, t)=\mathcal{U}_g(i-1, m, t).
\end{equation}

It is slightly more complicated to analyze the transitions of state $\mathcal{U}_g(i, m, t)$ when user $i$ is interested in tile $g$.  There are two transition possibilities for this state: 

(i) if the resolution level $m$ of $g$ is not transmitted, the recursive function is
\begin{equation}
\label{eq:interestNotTrans}
\mathcal{U}_g(i, m, t) = \mathcal{U}_g(i, m-1, t).
\end{equation}

(ii) If the resolution level $m$ is transmitted at link rate level $r_{i'}$ ($i' \leq i$), the recursive function is
{\fontsize{8}{9} \selectfont 
\begin{equation}
\label{eq:interestTrans}
\mathcal{U}_g(i, m, t) = \max\limits_{1\leq i' \leq i}\left \{\mathcal{U}_g(i'-1, m-1, t-\left \lceil{\frac{s_g^m}{r_i}}\right \rceil)+\sum\limits_{c=i'}^{i}u_{g,c}^m \right\}.
\end{equation}
}

The terminating conditions for the recursion and the corresponding value assignments are
\begin{align}
\mathcal{U}_g(i, m, t) &= -\infty, & \mbox{if}\ t < 0\text{ or } m < 0;  \nonumber \\
\mathcal{U}_g(0, m, t) &= 0,  & \ \mbox{if}\ t \geq 0\text{ and } m \geq 0.   \nonumber
\end{align}

We start the recursion from state $\mathcal{U}_g(n, M, t)$ with the given available time slots $t$, the highest resolution level $M$, and user $n$ with the highest link rate.  The recursion can be solved by applying Eqs. (\ref{eq:notInterested}), (\ref{eq:interestNotTrans}), and (\ref{eq:interestTrans}).  The transition complexity for Eqs.  (\ref{eq:notInterested}) and (\ref{eq:interestNotTrans}) are both $\mathcal{O}(1)$.    Eq. (\ref{eq:interestTrans}) enumerates the user link rate for every transmission to attain the optimal transition.  As a result, the transition complexity for Eq. (\ref{eq:interestTrans}) is $\mathcal{O}(n)$.  The overall computational complexity of our optimization algorithm is $\mathcal{O}(n^2tM)$, which grows quadratically with $n$.

\subsubsection{Virtual Clustering}
This section applies a clustering method to make our optimal algorithm scalable with $n$ (number of users).   Since Eq. (\ref{eq:interestTrans}) is the most time consuming operation, we will concentrate on analyzing this equation.

Assuming that a specific link rate $r_{i'}$ is used for transmitting resolution level $m$ in Eq. (\ref{eq:interestTrans}), all clients with no smaller than link rate $r_{i'}$ are able to receive this resolution level.  Instead of enumerating user $i'$, only the distinct link rates are required to be considered.  
As a consequence, we could cluster the users with identical link rate to a virtual user in the algorithm.  The clustering process is achieved by simply integrating the corresponding utility values.  Specifically, the utility of tile $g$ at resolution level $m$ for a virtual user at link rate $r$ could be defined as $\sum_{i=1}^n{u_{g, i}^m}$, where $r_{i} = r$.  

By clustering, the number of users $n$ is reduced to at most $N_r$, which is the maximum number of distinct link rates.  As the number of link rate levels is noticeably small ($8$ in $802.11a$~\cite{ieee2007ieee}), with user clustering, our algorithm scales with any number of users without considering the frame losses and retransmissions.

\subsection{Optimal Allocation for Multiple Tiles} \label{sec:optimalmtileallocation}
This section presents an algorithm that is able to achieve the maximum utility by optimally allocating resources over all $N_g$ tiles.  First, we extend the algorithm in Section \ref{sec:singleTileSchedule} to incorporate multiple tiles.  Next, we analyze the computational complexity of the algorithm and demonstrate its inefficiency.  Finally, we reduce the computational overhead of the algorithm to make it more efficient and practical for deployment.

Given time limit $t(g)$ for tile $g$, the optimal utility is $\mathcal{U}_g(n, M, t(g))$, which is calculated in Section \ref{sec:singleTileSchedule}.   The overall system utility is the integrated utility over all $N_g$ tiles, the optimization problem can be represented as
\begin{align}
\label{eq:overallSystemObjective}
\text{maximize} & \sum\limits_{g=1}^{N_g} \mathcal{U}_g(n, M, t(g)), \nonumber \\
\text{subject to } & \sum\limits_{g=1}^{N_g} { t(g) } \leq T.
\end{align}
From the formulas, we observe that optimization problem (\ref{eq:overallSystemObjective}) is to optimally distribute the total time slots $T$ to all tiles. 

Define function $\mathcal{U}(g, t)$ as the maximum utility achieved with tiles from $1$ to $g$ within time limit $t$.  Enumerating the allocated time slots $t'$ for transmissions of tile $g$ yields 
\begin{equation}
\label{eq:multipleTileSchedule}
\mathcal{U}(g, t) = \max\limits_{0 \leq t' \leq t}{\left \{  \mathcal{U}(g-1, t-t')+\mathcal{U}_g(n, M, t')  \right \}}.
\end{equation}
The maximum system utility is $\mathcal{U}(N_g, T)$.  This equation is employed by Li et al.~\cite{li2009scalable,li2010scalable} as well to incorporate the allocation of multiple multicast sessions into their optimal algorithm.

We now discuss the complexity of this multiple tiles allocation algorithm.  We precomputed all $\mathcal{U}_g(n, M, t)$,  where $1 \leq g \leq N_g$ and $0 \leq t \leq T$, the complexity is $\mathcal{O}(n^2TMN_g)$.  As shown in Eq. (\ref{eq:multipleTileSchedule}), the transition complexity for each state is $\mathcal{O}(T)$, the complexity of the recursion procedure to calculate $\mathcal{U}(N_g, T)$ is $\mathcal{O}(T^2N_g)$.  Combining the precomputing and the recursion complexity gives $\mathcal{O}(n^2TMN_g+T^2N_g)$ in total.

The parameters of $n$ (reduced to $N_r$), $M$, and $N_g$ are constants for a given video, so the computational cost dependents on $T$.  Assuming that the video frame rate is $25fps$, the slots available on average for a single frame is $40ms \approx 4444$ slots ($9\mu s$ per slot in $802.11a$).  When this value of $T$ is substituted into $\mathcal{O}(n^2TMN_g+T^2N_g)$, the overhead is clearly too large to be practical.  Therefore, it is essential to further reduce the computational complexity.

The key idea of reducing computational overhead is to trade space for algorithm running time.  Define the optimal utility function $\mathcal{U}^*(g, i, m, t)$ as
\begin{equation}
\label{eq:multipleTileStateDefine}
\mathcal{U}^*(g, i, m, t) = \max\limits_{0 \leq t' \leq t} {\left \{  \mathcal{U}(g-1, t-t') + \mathcal{U}_g(i, m, t') \right \}}.
\end{equation}

Same as the analysis for the allocation algorithm of a single tile, the category that each state $\mathcal{U}^*(g, i, m, t)$ falls into depends on whether user $i$ is interested in tile $g$.

If user $i$ is not interested in tile $g$, substituting Eq. (\ref{eq:notInterested}) into Eq. (\ref{eq:multipleTileStateDefine}) yields
\begin{align}
\label{eq:notInterestedMultipleTile}
\mathcal{U}^*(g, i, m, t) &= \max\limits_{0 \leq t' \leq t} {\left \{  \mathcal{U}(g-1, t-t') + \mathcal{U}_g(i-1, m, t') \right \}}  \nonumber \\
				   &= \mathcal{U}^*(g, i-1, m, t).
\end{align}

On the other hand, if user $i$ is interested in tile $g$, by substituting Eqs. (\ref{eq:interestNotTrans}) and (\ref{eq:interestTrans}) into Eq. (\ref{eq:multipleTileStateDefine}), we attain Eq. (\ref{eq:interestedMultipleTile}).

\begin{figure*}[t]
\normalsize
\setcounter{MYtempeqncnt}{\value{equation}}
\begin{align}
\label{eq:interestedMultipleTile}
\mathcal{U}^*(g, i, m, t) &= \max\limits_{0 \leq t' \leq t} {\left \{  \mathcal{U}(g-1, t-t') + \max{ \bigg[ \mathcal{U}_g(i, m-1, t'),  \max\limits_{1\leq i' \leq i}\bigg(\mathcal{U}_g(i'-1, m-1, t'-\frac{s_g^m}{r_{i'}})+\sum\limits_{c=i'}^{i}u_{g,c}^m \bigg) \bigg] }  \right\} }  \nonumber  \\
    &= \max {\left\{    \mathcal{U}^*(g, i, m-1, t), \max\limits_{1\leq i' \leq i} { \bigg[ \mathcal{U}^*(g, i'-1, m-1, t-\frac{s_g^m}{r_{i'}}) + \sum\limits_{c=i'}^{i}{u_{g,c}^m} \bigg] }   \right\} } 
\end{align}
\hrulefill
\vspace*{4pt}
\end{figure*}

The initial conditions and recursive transitions at boundaries for $\mathcal{U}^*(g, i, m, t)$ are
\begin{align}
\mathcal{U}^*(g, i, m, t) &= -\infty, & \mbox{if } t < 0\text{ or } m < 0;  \nonumber \\
\mathcal{U}^*(g, 0, m, t) &= \mathcal{U}^*(g-1, n, M, t), & \mbox{if } g \geq 1, t \geq 0, m \geq 0;  \nonumber \\
\mathcal{U}^*(0, i, m, t) &= 0, & \mbox{if } t \geq 0, m \geq 0.  \nonumber
\end{align}

The recursive Eqs. (\ref{eq:notInterestedMultipleTile}) and (\ref{eq:interestedMultipleTile}) clearly illustrate the procedure to solve the optimal multiple tiles allocation problem.  The maximum utility is $\mathcal{U}^*(N_g, n, M, T)$.

The transition Eq. (\ref{eq:notInterestedMultipleTile}) consumes $\mathcal{O}(1)$ complexity.  Eq. (\ref{eq:interestedMultipleTile}) enumerates user id $i'$ instead of time slots, thus the transition complexity is $\mathcal{O}(n)$.  Taking all transitions into consideration, we have a total computational complexity of $\mathcal{O}(n^2TMN_g)$.  Here, $n$ can be replaced by $N_r$ by clustering users according to the available link rate levels.  Compared with previous multiple tiles allocation algorithm, the computational complexity of current  algorithm is significantly reduced by a factor of $T$.  In the evaluation section, we will demonstrate the effectiveness of our optimal algorithm.

\section{Experimental Setup} \label{sec:implementation}
To evaluate our algorithm, we setup the following experimental system.

\begin{figure}[ht]
\centering
\includegraphics[width=0.4\textwidth]{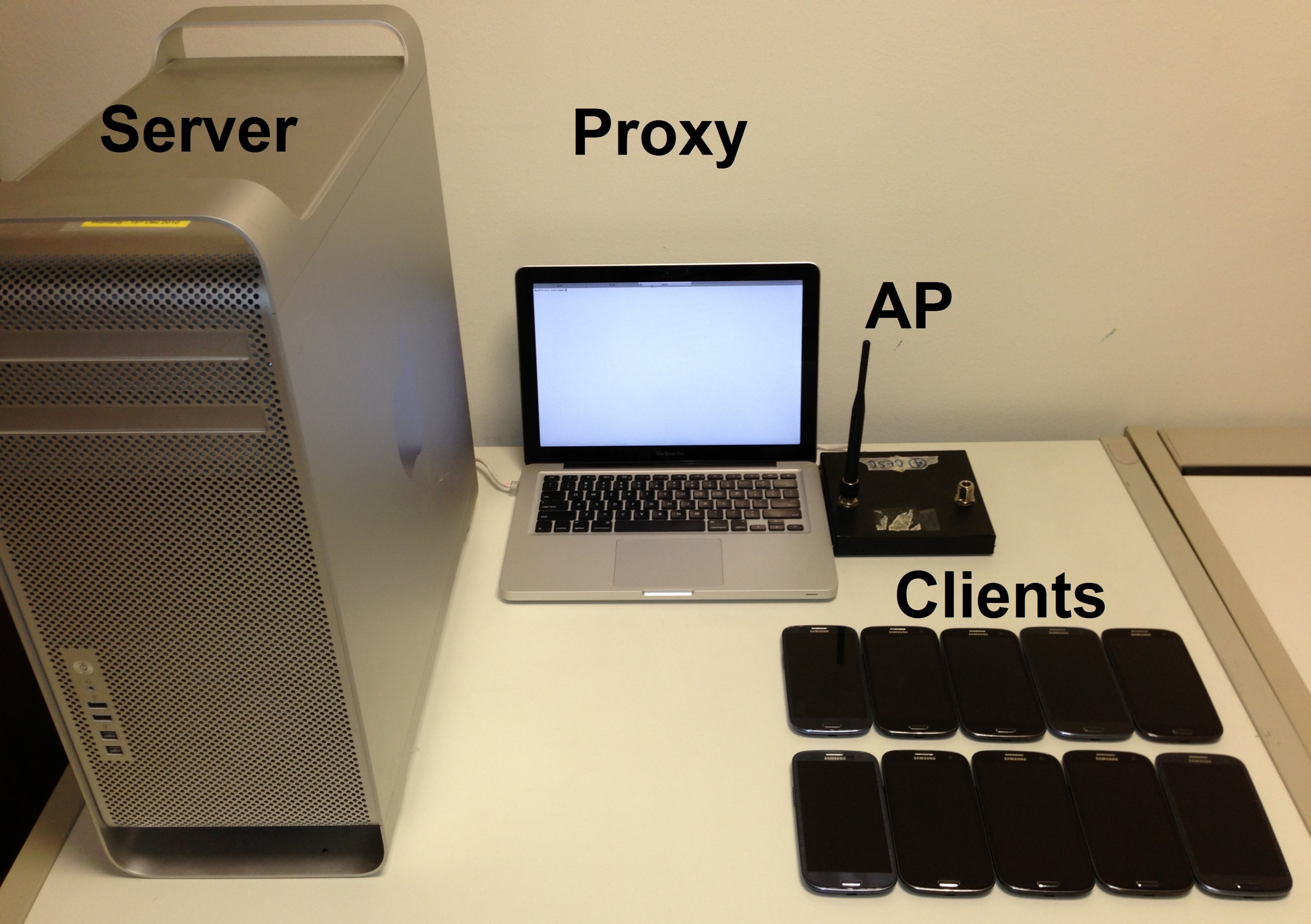}
\caption{System Setup.}
\label{fig:testbed}
\end{figure}

\subsection{System Setup}
Our system uses a zoomable video streaming server that runs on a Mac Pro with a 3.2GHz Quad-Core processor and 8GB memory.  The proxy runs on a MacBook with a 2.9GHz dual-core processor and 8GB memory. The video server, proxy, and WiFi AP used for multicast are all connected through wired Ethernet. The mobile devices, all Samsung Galaxy SIII, communicate with the AP using IEEE 802.11a operating at 5GHz.

The AP used supports two Complex IEEE 802.11abg adapters featuring the Atheros AR5414 chipset and runs OpenWRT Kamikaze 7.09 with kernel version 2.6.25.16. 
The driver of the wireless adapter used is MadWifi (version 0.9.4). To enable packet level rate assignment, we use the Click modular router \cite{kohler2000click} (version 1.6.0). For each video packet transmission, we extract the rate value that is specified by the proxy in the header of every video packet, then passes the assigned rate value to the MadWifi driver. The setup is shown in Figure \ref{fig:testbed}.



\subsection{Rate Adaptation}
As the WiFi SNR values on the mobile devices are not available, we use frame loss as a basis for rate adaptation \cite{bicket2005bit,wong2006robust,pefkianakis2013towards}.  In particular, we implement History-Aware Robust Rate Adaptation Algorithm (HA-RRAA) \cite{pefkianakis2013towards} that extends the work of RRAA \cite{wong2006robust}.  

RRAA uses two parameters, \emph{Maximum Tolerable loss} (MTL) and \emph{Opportunistic Rate Increase} (ORI), for rate adaptation.  The corresponding threshold for these parameters are denoted by $P_{MTL}$ and $P_{ORI}$, where $P_{ORI} < P_{MTL}$.  RRAA measures the frame loss rate $P$ over a period of \emph{Estimation Window} and
adapts the link rate as follows.  The rate decreases to next lower one if $P$ is greater than $P_{MTL}$.  If $P$ is smaller than $P_{ORI}$ the rate is increased to next higher one.  When $P$ is between $P_{MRL}$ and $P_{ORI}$, the current rate is retained.

To limit transmissions at the adjacent high loss rates, HA-RRAA is suggested \cite{pefkianakis2013towards}.  HA-RRAA exponentially increases the window size of next lower rate upon transmission failure of current rate ($P > P_{MTL}$) and reset the window size when transmissions of current rate are successful ($P < P_{MTL}$).  To be responsive to fast channel deterioration as RRAA, the algorithm additionally computes the loss over a \emph{small window}.  When the loss rate over this small window is greater than $P_{MTL}$, the current rate is directly moved to the next lower rate.

From our experiments, we observe that the HA-RRAA tuning mechanism may still result in the oscillation between two adjacent rates.  We slightly modify the algorithm so that the window size is halved instead of being reset when transmissions of the current rate is successful.  Furthermore, since we may broadcast packets at different rates under heterogeneous links, a client may receive packets sent at a rate higher than its current rate -- these packets serve as  ``free"  probes that prevent a client from increasing its
rate unnecessarily. As a result, our rate adaptation is stable and responsive.

For tractability, packet losses and frame retransmissions are not incorporated into our algorithm.  Therefore, conservative threshold parameters are used in our work.  In particular, we set $P_{MTL}=10\%$ and $P_{ORI} = 3\%$.  The minimum \textit{Estimation Window} size equals the interval between two consecutive allocation algorithm runs, this interval is also used as the \textit{small window} to maintain responsiveness.  

\subsection{Video Coding and Streaming}

In the evaluation, we do not need to play the video on the mobile devices and hence do not send actual video data. Instead, the following is done.
 
As depicted in Figure~\ref{fig:tilebroken}, each raw video frame from the test video is broken into $N_g$ tiles, and the tiles with same y-x are encoded using FFmpeg tool (version 1.2.1) with H264 codec at the server.  During our experiments, instead of transmitting the corresponding tiles from the test video, the server simply transmits the same number of arbitrary bits as the actual video tile.  The metadata containing the tile size, y-x position, resolution level, and the frame ID for identification, is embedded.  A client running on the mobile device extracts these fields from each received tile and periodically provide the reception bitmap to the server.  When the transmission is over, we gather the reception bitmaps from all the clients, and reconstruct the mixed-resolutions video frames with decoded tiles at the server side. Here, the lost tiles (indicated by bitmaps) in a group of pictures (GOP) are concealed by the default method in FFmpeg.

\section{Evaluation}
\label{sec:evaluation}

In this section, we present the evaluation results of our proposed optimal multicast algorithm through extensive experiments using up to 10 mobile devices.

\textbf{Compared Algorithms:}  We compare performance of optimal multicast against the following baseline schemes.  These schemes use HA-RRAA link adaptation as well. 

\emph{Adaptive Unicast (aUnicast)}: This scheme transmits packets using wireless unicast only.  To ensure the lowest quality (resolution level 1) is received by every user, the algorithm calculates the number of time slots required to transmit every tile at resolution level 1.   The algorithm then loops through each user, and if there is sufficient available time slot remaining, the resolution of the tiles transmitted to the user is replaced by the desired resolution level.  The loop terminates when the requests from all users are satisfied or the remaining time slots are insufficient for any user.


\emph{Adaptive Multicast (aMulticast)}: Similar to aUnicast, the lowest resolution level 1 is guaranteed for each user and the remaining available slots are utilized to upgrade the resolution level tile by tile.  As in DirCast \cite{chandra2009dircast}, the assigned link rate for a particular tile is the lowest supported link rate among all interested users.  As multicast is used, at most one multicast transmission is required for any tile.

\emph{Approximation}: We apply the approximation method in~\cite{li2010scalable} to our maximization problem, where the utility slots instead of the time slots is used as a state dimension in the dynamic programming.  The approximation factor bound of this approach is $1 - \varepsilon N_g$.  A better approximation factor is obtained with a finer-grained utility unit (a smaller $\varepsilon$).  As the computational complexity of the approximation algorithm grows quadratically with the number of utility units, the finer-grained utility unit significantly increases the computational complexity.  In our experiment, the same $\varepsilon = 0.2$ is used, the running time is close to our optimal multicast.

In our paper, all the above algorithms collect the RoI requests and run the allocation algorithm every 2 seconds.  The average running time of our optimal algorithm is 49.18ms, which only incurs 2.5\% overhead.

We measure the peak signal-to-noise ratio (PSNR), a standard metric for measuring the video quality, and \emph{goodput} of the system to compare the performance of the algorithms.


\textbf{Video Setup:}  We evaluate the algorithms using two standard HD (1920x1080p) test video files, \emph{controlled-burn} (dense motion) and \emph{rush-hour} (low motion)\footnote{Available at \url{http://media.xiph.org/video/derf/}}.  Table \ref{tab:resolutionsize} presents the video configurations and data rates.

\begin{table} [ht]
\renewcommand{\arraystretch}{1.2}
\caption{The data rate ($M\MakeLowercase{bps}$) of different resolution levels.}
\label{tab:resolutionsize}
\centering
\begin{threeparttable} [b]
\begin{tabular}{ | c | c | c || c | c | c | } 
\cline{1-6}
{level} & resolution size & \# tiles &  low rate\tnote{a} & medium rate\tnote{b} & high rate\tnote{c}  \\ \cline{1-6}
5 &1920$\times$1080 & 16$\times$9 & 6.2 & 10.9 & 20.2  \\ \cline{1-6}
4 &1600$\times$900  & 16$\times$9 & 4.5 & 6.6 & 11.1 \\ \cline{1-6} 
3 &1280$\times$720 & 16$\times$9 & 3.2 & 4.6 & 8.4  \\ \cline{1-6}
2 &960$\times$540 & 16$\times$9 & 2.2  & 2.9 & 5.0  \\ \cline{1-6}
1 &640$\times$360 & 16$\times$9 & 1.2 & 1.5 & 2.5 \\ \cline{1-6}
\end{tabular} 
\begin{tablenotes} 
\item[a] \emph{Rush-hour}, compressed using FFmpeg with parameter $qp=25$.
\item[b] \emph{Controlled-burn}, compressed with $qp=25$.
\item[c] \emph{Controlled-burn}, compressed with $qp=22$.
\end{tablenotes} 
\end{threeparttable} 
\end{table} 

\textbf{Wireless Channels:} We place the mobile devices at different locations and distances from the AP, to vary the channel conditions between the mobile devices and the AP.  Table \ref{tab:linkquality} shows the minimum, maximum, and average achieved link rates when there are up to 10 mobile devices. 

\begin{table} [ht]
\renewcommand{\arraystretch}{1.2}
\caption{The achieved link rates of mobile users ($M\MakeLowercase{bps}$).}
\label{tab:linkquality}
\centering
\begin{threeparttable} [b]
\begin{tabular}{ | c || c | c | c | } 
\cline{1-4}
\# users & min rate &  max rate & average rate \\ \cline{1-4}
1 &6 & 6 & 6 \\ \cline{1-4}
3 &6 & 36  & 20.0  \\ \cline{1-4}
5 &6& 36 & 21.6   \\ \cline{1-4}
8 &6  & 36 & 22.5  \\ \cline{1-4} 
10 &6 & 36 & 21.0  \\ \cline{1-4}
\end{tabular} 
\end{threeparttable} 
\end{table} 


\textbf{RoI Variation:} User requests and RoI used in the evaluations are based on the real interaction logs from 10 users who have used zoomable video system \cite{quang2010supporting}.

\subsection{Baseline Comparison}
The average PSNR with error bars (standard deviation) across different users streaming at medium video rate are depicted in Figure \ref{fig:mq_mr_psnr}. The corresponding achieved average goodput is present in Table \ref{tab:mq_mr_goodput}.  As the unicast scheme cannot fit the lowest resolution level requirement for more than 5 clients, no data point is presented in this range in the results.  From the results, we can draw the following observations:

\begin{figure*}[ht]
\centering
\begin{minipage}[b]{0.32\linewidth}
\centering
\includegraphics[width=\textwidth]{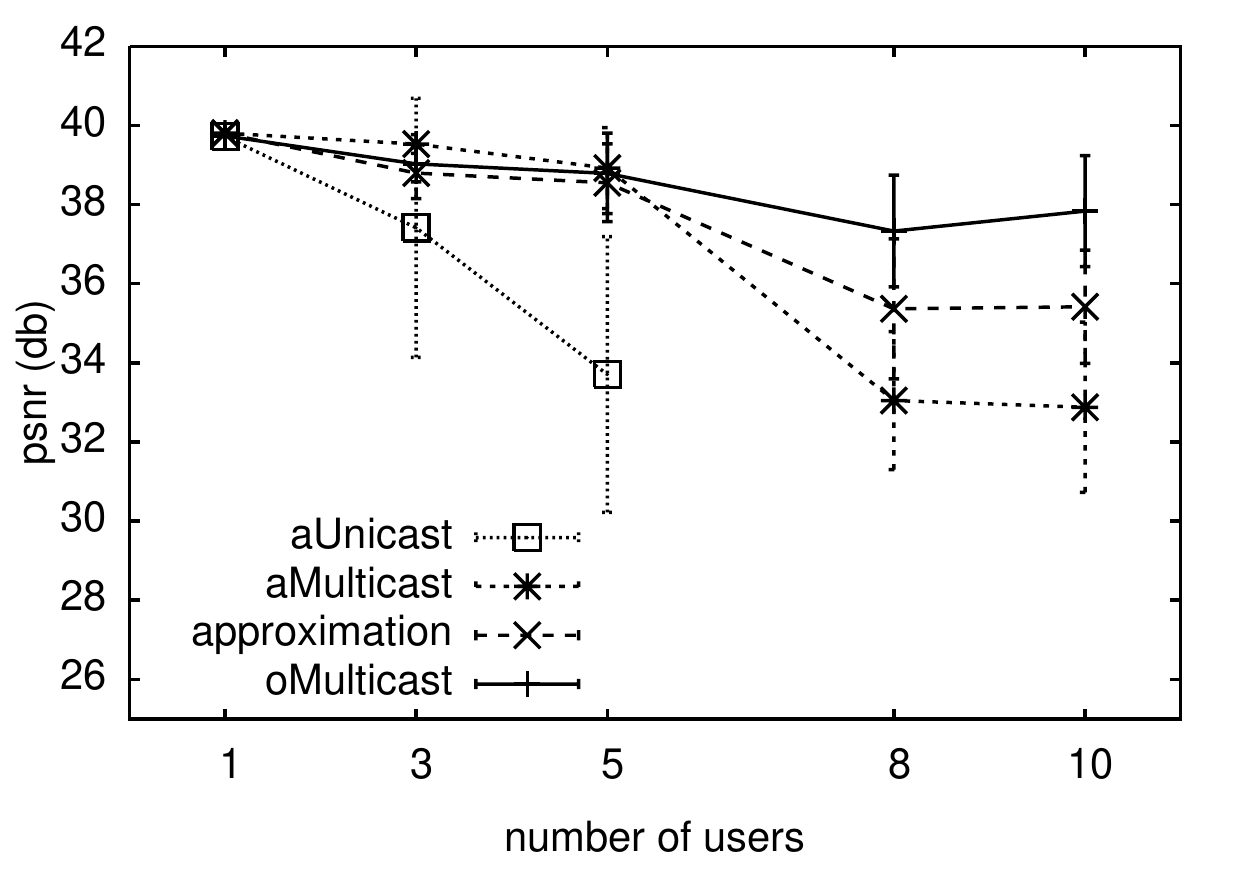}
\caption{Average PSNR with medium video rate.}
\label{fig:mq_mr_psnr}
\end{minipage}
\begin{minipage}[b]{0.32\linewidth}
\centering
\includegraphics[width=\textwidth]{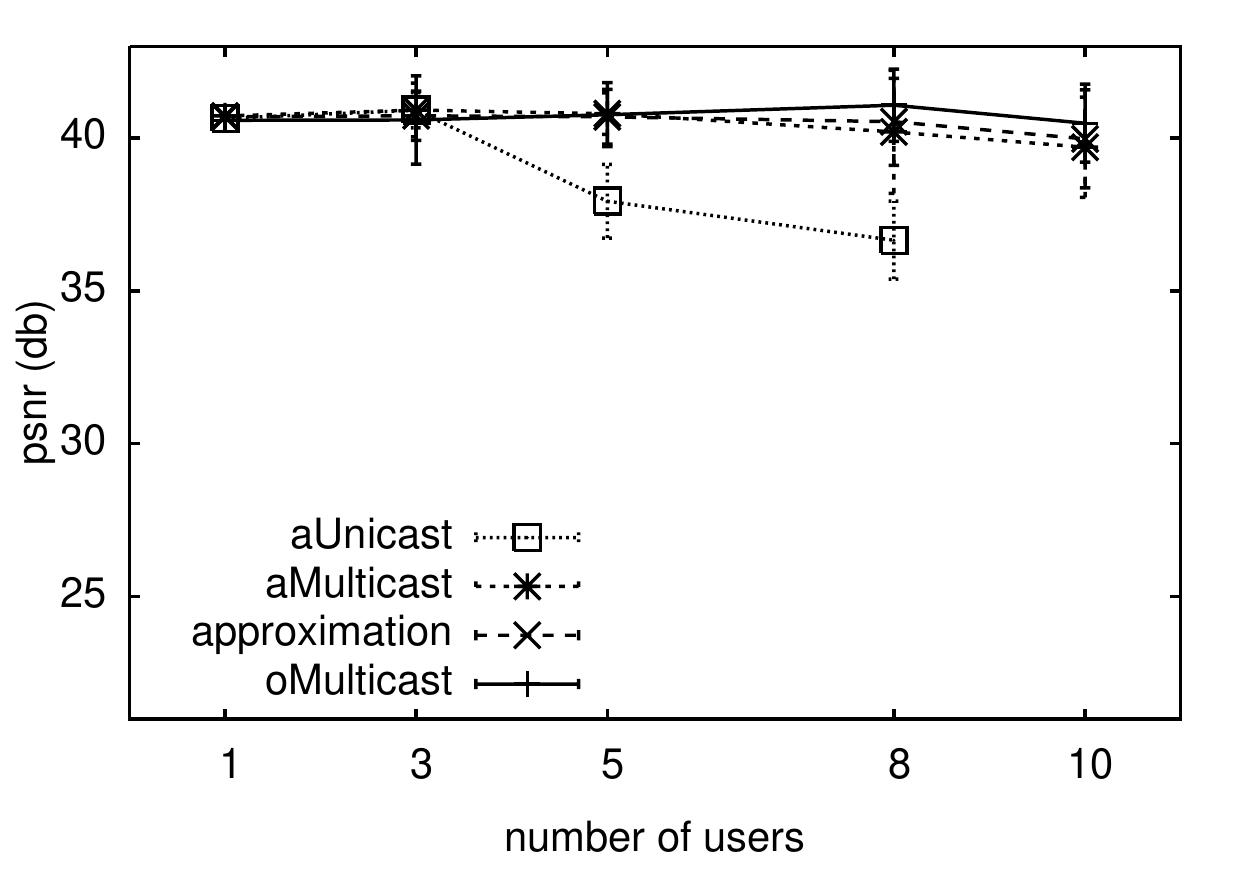}
\caption{Average PSNR with low video rate.}
\label{fig:mq_lr_psnr}
\end{minipage}
\begin{minipage}[b]{0.32\linewidth}
\centering
\includegraphics[width=\textwidth]{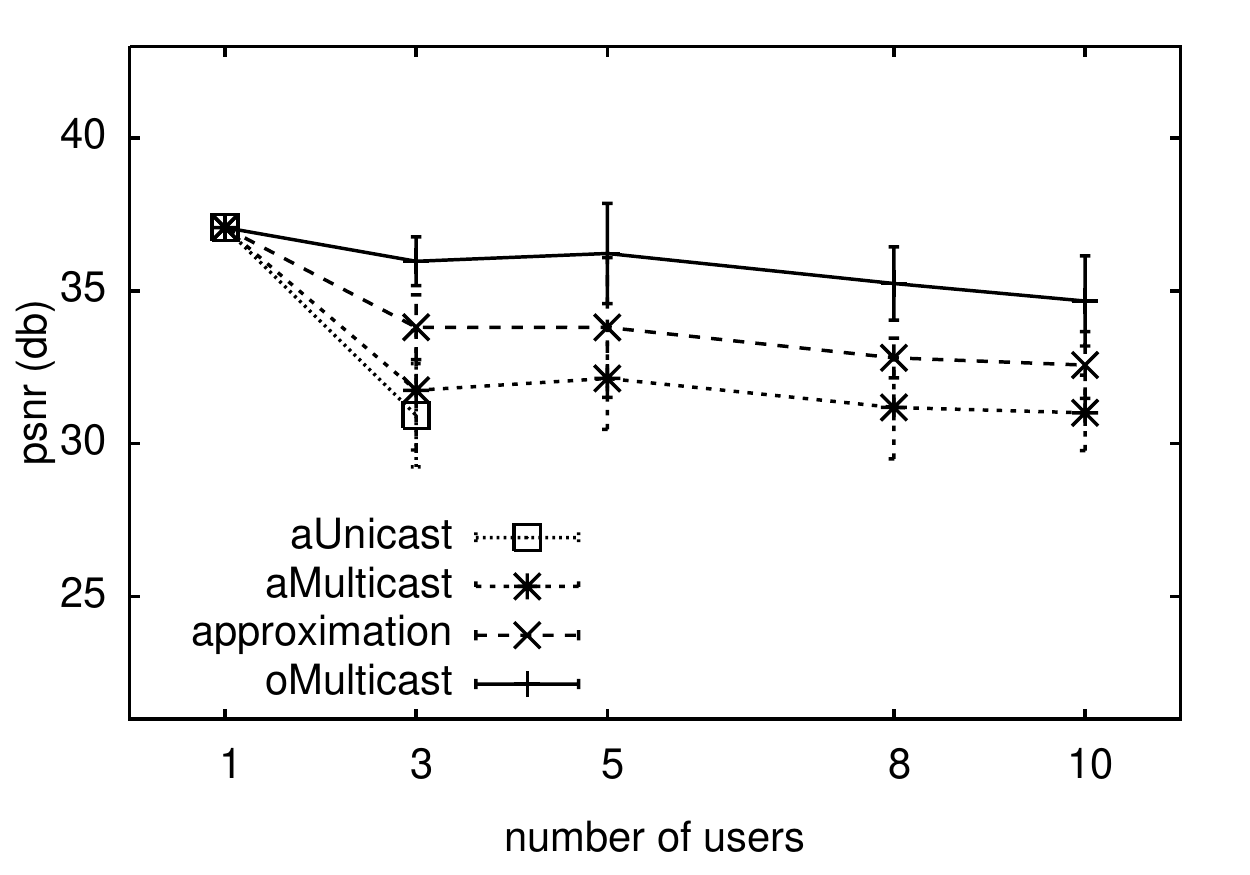}
\caption{Average PSNR with high video rate.}
\label{fig:mq_hr_psnr}
\end{minipage}
\vspace*{4pt}
\end{figure*}

(i) \emph{PSNR gains}.  The multicast algorithms are able to satisfy up to 5 users requests without notable PSNR degradation. On the other hand, the video quality with unicast dramatically decreases beyond 3 users, and only up to 5 users can be supported by adaptive unicast.  With more than 5 users, all three multicast schemes experience some PSNR loss. The optimal multicast, however, considerably outperforms approximation and adaptive multicast under heavy load, with the improvements of about 3dB and 5dB in PSNR, respectively. 

(ii) \emph{Goodput gains}.  Due to zooming, the demands between different clients are not identical. Hence, the trend in average goodput does not strictly follows that of video quality (Table \ref{tab:mq_mr_goodput}). 
As predicted from Figure \ref{fig:mq_mr_psnr} and Table \ref{tab:mq_mr_goodput}, the multicast algorithms outperform unicast when there are more than 3 users in terms of both PSNR and goodput. 
When there are more than 5 users, the improvements of optimal multicast over approximation and adaptive multicast with 10 users are 19\% and 34\%, respectively.

\begin{table} [t]
\renewcommand{\arraystretch}{1.2}
\caption{Average goodput ($M\MakeLowercase{bps}$) achieved with heterogeneous link qualities at medium video rate}
\label{tab:mq_mr_goodput}
\centering
\begin{threeparttable} [b]
\begin{tabular}{ | c || c | c | c | c | } 
\cline{1-5}
{\# users} & aUnicast & aMulticast & approximation & oMulticast  \\ \cline{1-5}
1 & 3.83 & 3.79 & 3.81 & 3.82 \\ \cline{1-5}
3 & 2.95 & 3.45 & 3.46 & 3.41  \\ \cline{1-5}
5 & 1.8 & 3.07 & 3.05 & 3.07  \\ \cline{1-5}
8  & $\backslash$ & 2.1 & 2.27 & 2.56 \\ \cline{1-5} 
10 & $\backslash$ & 1.99 & 2.25 & 2.67  \\ \cline{1-5}
\end{tabular} 
\end{threeparttable} 
\end{table} 

(iii) \emph{Fairness gains}.  The error bars in Figure \ref{fig:mq_mr_psnr} indicate that our optimal multicast achieves the best fairness among all algorithms, due to adaptive utility assignment (Section \ref{sec:aUtilityAssign}) in our algorithm.  Although a similar allocation method is used by adaptive multicast and adaptive unicast, they performs remarkably different in terms of fairness. While multicast transmission can benefit multiple users, unicast transmission does not, which may lead to less fairness among the users.


\subsection{Impact of Video Rate}
To evaluate the impact of video data rate (and thus the traffic load), we repeat the experiments using a different video with a lower rate and the same video encoded with a higher rate.  We generate \emph{low rate} and \emph{high rate} videos in addition to the previously used \emph{medium rate}.  The configurations are detailed in Table \ref{tab:resolutionsize}. 
Wireless link quality settings are the same to the previous section.
Figure \ref{fig:mq_lr_psnr} and Figure \ref{fig:mq_hr_psnr} depict the average achieved PSNR for low rate and high rate videos, respectively.

Figure \ref{fig:mq_lr_psnr} demonstrates that all four algorithms perform better with lighter workload as expected. Specifically, the multicast algorithms scale up to 10 users without significant quality degradation, and the unicast scheme is able to support more clients.

For higher traffic load, all algorithms perform worst.   Compared with other schemes, our optimal algorithm, however, still provide relatively fair quality under the higher load.  In general, if a client does not induce lower link rate or request higher resolution level, no additional multicast traffic will be introduced.  Thus, the video qualities are only slightly reduced even as more clients are added to the multicast sessions. 

\subsection{Impact of RoI Similarity}
Intuitively, larger amount of RoI overlapping increases the relative performance gap of multicast over unicast.
The impact of RoI overlapping is evaluated in this section. In order to control the amount of overlap, we do not use collected traces to simulate RoI variation. 
Instead, we manually vary the RoI sizes and positions so that they can change in a uniform and controlled manner. Here, the RoI sizes and the request resolution levels of all clients are identical. We vary the positions of RoI to generate different similarity.
\begin{figure}[ht]
\centering
\includegraphics[width=0.4\textwidth]{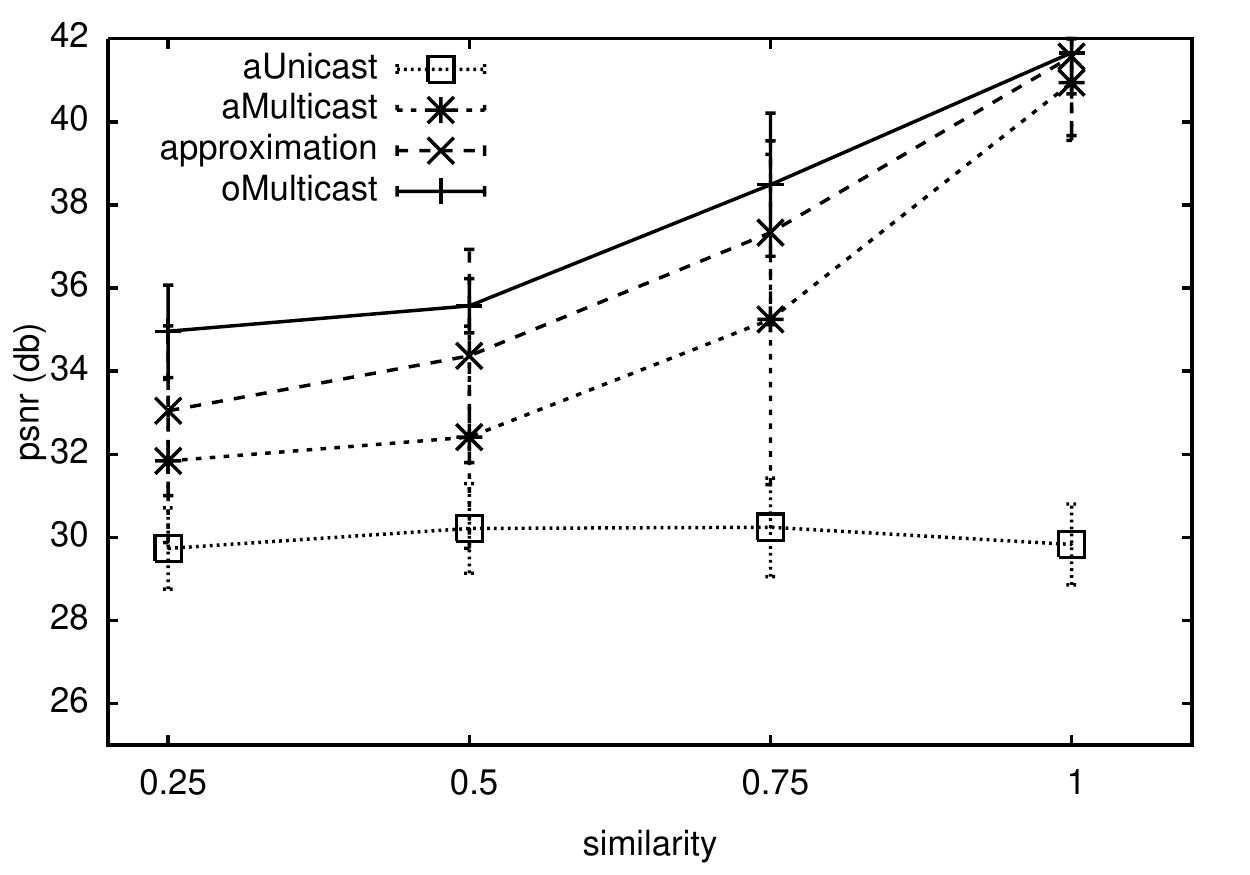}
\caption{Average PSNR with different similarity.}
\label{fig:similarity_psnr}
\end{figure}

To measure the degree of overlapping, we first define the \textit{popularity} of a tile $g$, $p_g$ as the fraction of the number of users interested in it.  The degree of overlapping for user $i$ is then the total popularity of all tiles in $\mathbf{G}(i)$, excluding the tiles only interested by user $i$, divided by the number of tiles in $\mathbf{G}(i)$.  We then define \textit{similarity} as the average overlapping degree across all users.  We present how PSNR changes with different similarity, for 8 users, in Figure \ref{fig:similarity_psnr}.

The relatively stable performance in terms of video quality 
shows that the unicast scheme is not affected by the amount of RoI overlap. As expected, the improvement of multicast over unicast increases with the increasing RoI similarity. When the RoIs are identical (all users want the same regions), the improvement is about 12dB in PSNR.
Interestingly, with increased similarity value, the PSNR quantities of three multicast algorithms converge to an identical point.  Such convergence is caused by both the decrease in traffic demand and the fact that the same data is requested.

\subsection{Client Mobility}
\begin{figure}[h]
\centering
\includegraphics[width=0.4\textwidth]{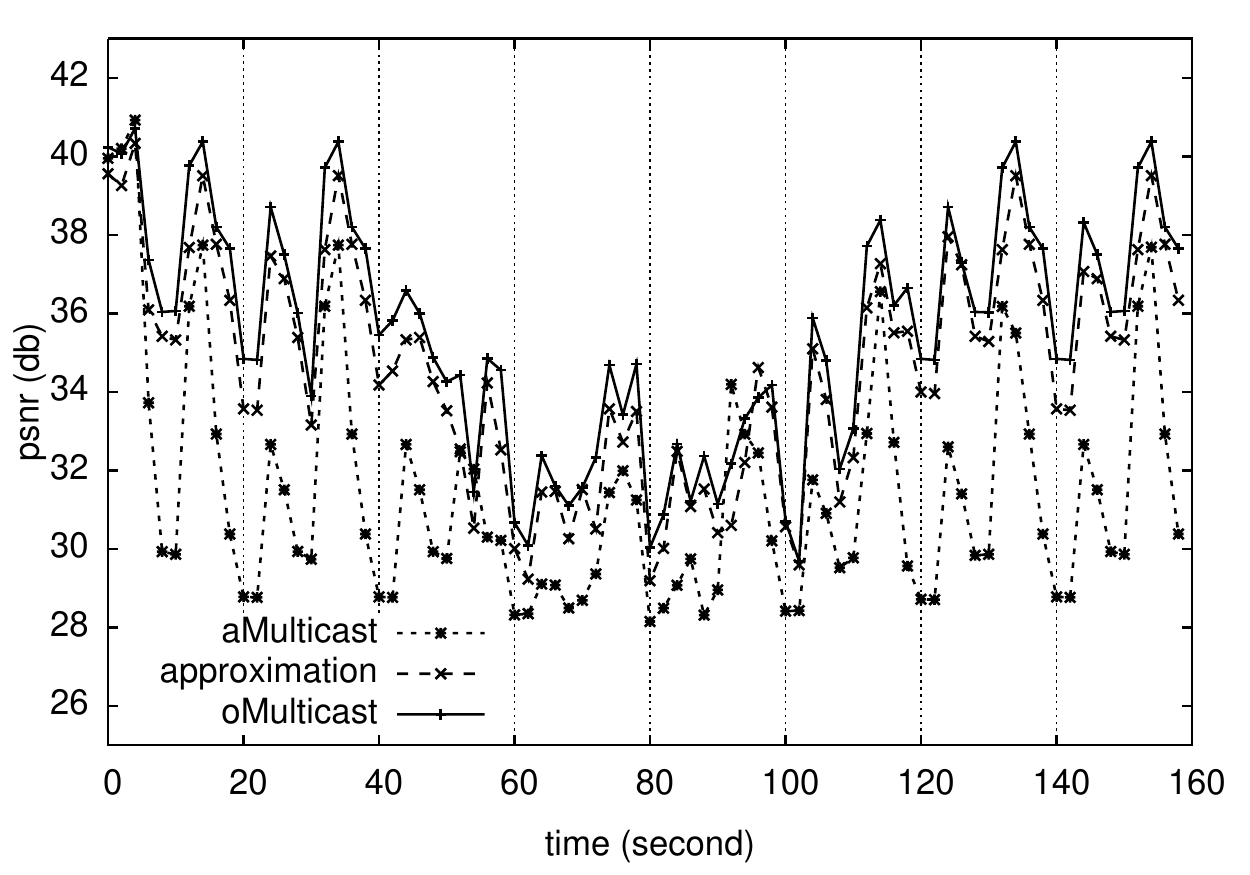}
\caption{Average PSNR of the mobile client.}
\label{fig:mobility}
\end{figure}

The previous sections demonstrate the effectiveness of our optimal multicast algorithm with stationary clients.  In this section, we evaluate the performance of our optimal algorithm with client mobility.  In particular, we keep two clients static, the obtained link rates for them are 6Mbps and 36Mbps.  One additional mobile client starts from a location close to the AP, moves away from it, and then moves back.  Figure \ref{fig:mobility} plots the average PSNR of the mobile client for every two seconds.  The movement period is from 40s to 120s.

In the experiment, the high rate video is used and a segment of 20s is played repeatedly.  Although, the RoI of each user and the allocations are fixed under static condition, the PSNR of different frames are different.  This disparity is due to the fact that sensitivity of different frames with mixed resolution tiles are different.  The same trend of PSNR variations under static conditions can be observed between different playbacks.

From the figure, we observe that our optimal algorithm consistently outperforms two baseline algorithms.  The average enhancements of our optimal multicast over approximation and adaptive multicast are about 1dB and 4.5dB, respectively.   Moreover, our algorithm can quickly adapt to the link rate and the video quality returns quickly to the level similar to the static period after the movement (at 120 second).

\section{Conclusion} \label{sec:conclusion}
We have developed and implemented an efficient algorithm for multicasting mixed resolution tiles to heterogeneous users, for interactive video applications that support zoom and pan.  Our algorithm optimizes the total utility of all clients and achieves significant improvements in video quality: up to a 3dB improvement over approximation multicast approach, 6dB improvement over an adaptive multicast scheme, and 12dB improvement over adaptive unicast scheme in our experiment settings.  Additionally, our approach can be directly applied to design an optimal allocation algorithm for a general multi-sessions video multicast.  In the future, we shall extend this work to the scenarios with multiple access points (APs), where the AP association mechanism could be exploited to further enhance the multicast performance.

\balance

\end{document}